\documentclass[12pt]{article}

\usepackage{color}
\usepackage{scicite}
\usepackage{epsfig}
\usepackage{subfigure}
\usepackage{amssymb}
\usepackage{amsmath}
\usepackage[dvipsnames]{xcolor}
\usepackage{xspace}
\usepackage{hyperref}
\usepackage{graphicx}

\newcommand{\aap}{    {\it Astron. Astrophys.}}
\newcommand{\aaps}{   {\it Astron. Astrophys. Suppl. Ser.}}
\newcommand{\aapr}{   {\it Astron. Astrophys. Rev.}} 
\newcommand{\aj}{     {\it Astron. J.}} 
\newcommand{\apj}{    {\it Astrophys. J.}}
\newcommand{\apjl}{   {\it Astrophys. J.}}  
\newcommand{\apss}{   {\it Astrophys. Space Sci.}}
\newcommand{\mnras}{  {\it Mon. Not. R. Astron. Soc.}}

\newcommand{\solphys}{{\it Sol. Phys.}}
\newcommand{\araa}{    {\it Annu. Rev. Astron. Astrophys.}}
\newcommand{\icarus}{    {\it Icarus}}

\newcommand{\arthur}{HD~187160\xspace}

\def\teff{T_{\text{eff}}}

\def\otz{\Omega_0} 
\def\oto{\Omega_1}

\newenvironment{sciabstract}{%
\begin{quote} \bf}
{\end{quote}}

\title{Asteroseismic detection of latitudinal differential rotation in 13 Sun-like stars}
\author{O. Benomar$^{1\ast}$, M. Bazot$^{1}$, M.B. Nielsen$^{1}$, L. Gizon$^{2,3,1}$, \\
  T. Sekii$^{4}$, M. Takata$^{5}$, H. Hotta$^{6}$, S. Hanasoge$^{7,8}$, \\ 
K.R. Sreenivasan$^{1,8}$, J. Christensen-Dalsgaard$^{9}$ \\
\\
\normalsize{$^{1}$Center for Space Science, New York University Abu Dhabi, UAE}  \\
\normalsize{$^{2}$Max-Planck-Institut f\"ur Sonnensystemforschung, 37077 G\"ottingen, Germany} \\
\normalsize{$^{3}$Institut f\"ur Astrophysik, Georg-August-Universit\"at G\"ottingen, 37077 G\"ottingen, Germany} \\
\normalsize{$^{4}$National Astronomical Observatory of Japan, Mitaka, Tokyo 181-8588, Japan} \\
\normalsize{$^{5}$Department of Astronomy, The University of Tokyo, Bunkyo-ku, Tokyo 113-0033, Japan} \\
\normalsize{$^{6}$Department of Physics, Graduate School of Science, Chiba university, Chiba 263-8522, Japan} \\
\normalsize{$^{7}$Tata Institute of Fundamental Research, Mumbai, India 400005} \\
\normalsize{$^{8}$New York University, NY 10012, USA} \\
\normalsize{$^{9}$Stellar Astrophysics Centre, Department of Physics and Astronomy, Aarhus University, Denmark} \\ 
\\
\normalsize{$^\ast$To whom correspondence should be addressed; E-mail:  othman.benomar@nyu.edu.}
}

\date{}

\begin{document} 


\baselineskip24pt


\maketitle 


\begin{sciabstract} 
The differentially rotating outer layers of stars are thought to play a role in driving their magnetic activity, but the underlying mechanisms that generate and sustain differential rotation are poorly understood. 
We report the measurement of latitudinal differential rotation in the convection zones of 40 Sun-like stars using asteroseismology. 
For the most significant detections, the stars' equators rotate approximately twice as fast as their mid-latitudes. The latitudinal shear inferred from asteroseismology  
is much larger than predictions from numerical simulations.
  \end{sciabstract}

Analysis of acoustic oscillations visible on the Sun's surface using helioseismology has been critical for constraining its rotation profile. Helioseismology has revealed that the rotation rate of the Sun's convection zone decreases with
latitude \cite{Thompson2003,Schou1998}. 
This latitudinal differential rotation has a magnitude of $11\%$ of the average rate from equator to mid-latitudes ($\approx 45^\circ$ latitude) and $30\%$ between equator and the poles. At the base of the convection zone, the Sun transitions to solid-body rotation. How such a rotation profile is established and maintained is still poorly understood. However, it is likely that differential rotation plays a role in sustaining the solar magnetic field through a dynamo mechanism \cite{Charbonneau2010,Ossendrijver2003,Parker1955}.

So far, little is known about the latitudinal differential rotation in other stars, and classical methods for investigating it are primarily sensitive to the near-surface layers. Most studies rely on photometric variability from starspots at different latitudes \cite{Olah2009}, 
or use Doppler imaging to track magnetic features at the surface and their migration in latitude \cite{Barnes2005,Donati1997}. Another approach involves studying the Fourier transform of the spectroscopic line profiles \cite{Reiners2001,Gray1977}. 

Asteroseismology provides an opportunity to probe rotation inside stars, including Sun-like pulsators \cite{Benomar2015,Gizon2013,Appourchaux2008}, since 
it studies the resonant frequencies of waves within the body of the star. 
Among these are acoustic waves which travel at various depths and have a varying sensitivity to rotation with latitude.
These accoustic waves can be used to infer the star's internal rotation both in radius and latitude \cite{Lund2014,Aerts2010,Gizon2004,Gizon03}. 

The NASA {\it Kepler} spacecraft has provided high-precision, long-duration photometric time series for many stars, which is necessary for the study of differential rotation of Sun-like stars with asteroseismology.

Oscillations of Sun-like stars are driven by the stochastic convective motion of material in the outer envelope of the stars. 
Each mode of oscillation is identified by the overtone number $n$ and spherical harmonic functions of angular degree $l$ and azimuthal order $m$. Modes with the same $n$ and $l$ but different $m$ appear as multiplets of $2l+1$ components.
The splitting of modes in a multiplet provides information  about the rotation profile and on the  physical processes acting within the star (such as flows, internal structure, tidal forces).

An efficient means for quantifying the impact of rotation on the split frequencies of acoustic waves is through the use of Clebsch-Gordan \emph{a}-coefficients  $a_{1}, a_{2}, a_{3}, ...$ \cite{Schou1994}. 
This decomposition has been used extensively in helioseismology but not in the analysis of other stars. 
For stars other than the Sun, only low-degree modes ($l \leq 3$) can be observed and so only the coefficients $a_1$, $a_2$, and $a_3$ can be determined. The coefficient $a_1$ is an average of the rotation rate. The $a_2$ coefficient is related to the asphericity \cite{Gizon2016}, while $a_3$ is a measure of the latitudinal differential rotation \cite{Gizon2004}. The coefficient $a_3$ is positive when the pole rotates slower than the equator, that is if the star exhibits a solar-like rotation profile (Figure S1). Conversely, $a_3$ is negative for anti-solar rotation profiles. 
Simulations tend to show that fast rotation causes solar-like rotation while slower rotation rates lead to anti-solar rotation \cite{Gastine2013,Hotta2015,Fan2014}
although simulations found an anti-solar rotation, at solar rotation rates \cite{Hotta2015,Fan2014}. This is evidently incompatible with solar observations and constraints from other stars are necessary to resolve this issue.
 
Pulsations appear as Lorentzian profiles in a power spectrum of the star's photometric time series \cite{benomar2017science_supmat}.
The analysis of the pulsations is performed by fitting a model to the power spectrum, using an MCMC sampling technique \cite{Benomar2015}.
The contribution of $a_{1}$ to mode frequencies is easily measured, but the absolute value of $a_{3}$ is approximately two orders of magnitude smaller than $a_{1}$ so it is only possible to measure it using multi-year observations from space instruments such as {\it Kepler}.

\begin{figure*}[t]
  \begin{center}	\subfigure{\epsfig{figure=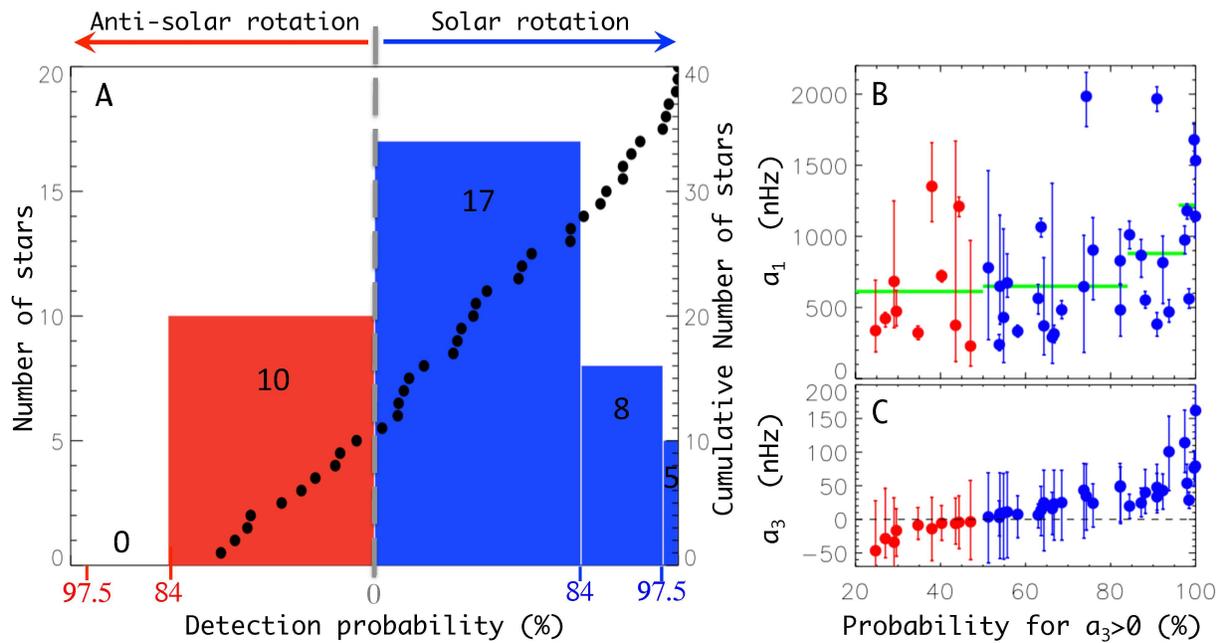, angle=-90, width=16cm}}
  \end{center}
\caption{\textbf{Detection of solar rotation and anti-solar rotation.} \textbf{(A)} Histogram of the detection significance for solar-like versus anti-solar rotation using {\it Kepler} asteroseismic lightcurves (colored bars). The cumulative distribution is also shown (black dots). All stars with conclusive detections of $a_3 \neq 0$ ($>84\%$ significance) have solar-like rotation ($a_3 > 0$). \textbf{(B)} Average internal rotation as measured by $a_1$. Stars with high detection significance rotate faster than those with low significance. \textbf{(C)} the latitudinal differential rotation coefficient $a_3$. }  
\label{fig:detect_significance}
\end{figure*}

We perform a combined measurement of the $a_1$ and $a_3$ coefficients for 40 stars of mass between  $0.9M_{\odot}$ and $1.5M_{\odot}$, observed as part of the {\it Kepler} LEGACY sample \cite{Lund2017,Aguirre2017}. 
By integrating the probability density functions of $a_3$ determined by fitting the power spectra, we compute the detection significance for either solar or anti-solar latitudinal differential rotation (Figure \ref{fig:detect_significance} and Table S1). Applying a detection threshold at a probability of  $84\%$ (i.e. excluding $a_3=0$ with a detection significance $>1\sigma$ for a Gaussian)  we find that none of the stars unambiguously show anti-solar rotation while $32\%$ (13 stars) show significant solar-like rotation. Five stars have a detection probability of more than $97.5\%$ (or a significance $>2\sigma$). 
The excess of solar-like rotators may be due to an observational limitation. Antisolar differential rotation is theoretically expected \cite{Gastine2013,Featherstone2015} for slow rotators, in which $a_1$ and $a_3$ are difficult to measure. Our sample consists predominantly of fast rotators, and so we may only be sensitive to stars with solar-like rotation profiles. 
Our most statistically significant detections have relatively high rotation rates compared to those below the detection threshold.

The Rossby number is an indicator of the influence of rotation on a fluid compared to convective motion. Current theory \cite{Gastine2013} predicts that solar-like rotators should have a Rossby number less than unity. For the majority of our significant detections, we find a Rossby number less than $0.8$ (see Figure S2). 

We now seek to estimate the rotational shear between the equator and higher latitudes. For this, we select two representative stars  (HD~173701 and HD~187160) in terms of $a_1$, $a_3$ and internal structure (i.e. thickness of convection zone). 
Their power spectra are shown in Figure S3-4 with the best-fiting model. HD~173701 is a cool dwarf of spectral type G8 (the Sun is of spectral type G2), with a mass $0.974 \pm 0.029\,M_\odot$ and age $4.69\pm 0.44$ Gyrs \cite{Lund2017}. It is known to be very active \cite{Kiefer2017}, with irregular variability on timescales that range from 8 to 40 days, likely caused by starspots. The probability density function deduced from the fit indicates that $a_1=574.6 \pm 82.0$~nHz and $a_3=28.4 \pm 12.5$~nHz (Figure S5).
\arthur is of spectral type F9 and has a mass and age of $1.09 \pm 0.03\,M_\odot$ and $3.28\pm0.16$ Gyrs. Thus, \arthur is similar to a younger version of the Sun. It has a visible magnitude of $7.4$ and a faint, cool spectroscopic binary companion of magnitude $8.7$ \cite{Mason2001}. The average internal rotation period is $\sim9$ days ($a_1 = 1185.0 \pm 51.7$~nHz).
The light curve of \arthur shows variability at two distinct time scales, one at $\approx 9.4$ days and another at $\approx 17.5$ days. We attribute this variability to starspots on the surfaces of \arthur and its fainter non-interacting companion, respectively.
We measure $a_3= 53.7 \pm 25.0$~nHz (Figure S6). 

We have evaluated the potential systematic errors associated with the mode-fitting methodology using simulated power spectra for HD~173701 and HD~187160 \cite{benomar2017science_supmat}. These systematic errors are small, meaning $a_{3}$ in both cases is positive with a probability higher than $97.5\%$. This confirms that these stars have poles that rotate slower than the equator. 

To obtain a latitudinal rotation profile we use an inversion of the $a$-coefficients \cite{Gizon2004}. Because we only measure $a_1$ and $a_3$, we are restricted to a two-parameter rotation model for this.
We assume a solar-like rotation profile $\Omega(\theta) = \Omega_{0} + 3 \Omega_1 (5 \cos^2 \theta - 1)/2$ in the convection zone, where $\theta$ is the co-latitude (complementary angle of the latitude such that $\theta=0$ is the pole and $\theta=90^\circ$ is the equator).
The average internal rotation rate is given by $\Omega_0$ while the term in $\Omega_1 \propto \Omega_{\rm pole} - \Omega_{\rm eq}$ measures the contrast in rotation (latitudinal shear) from the equator to the pole in the convection zone. In the equation, $\Omega_{\rm pole}$ and $\Omega_{\rm eq}$ are the pole and equator rotation rate, respectively. The convection zone extends from the surface down to $0.687 R_\star$ for HD~173701 and to $0.785 R_\star$ for HD~187160, where $R_\star$ is the stellar radius. 
Due to its relatively fast rotation, \arthur was investigated for possible variations of $a_1$ with the overtone $n$ (Figure S7). 
We did not find evidence of significant variation.  
Studies of radial differential rotation of low-mass main-sequence stars indicate that variation in $a_1$ should not exceed ~30\% \cite{Nielsen2017}, and we therefore do not expect this to have an effect on the inversion.

\begin{figure*}[t]
  \begin{center}
  	\subfigure{\epsfig{figure=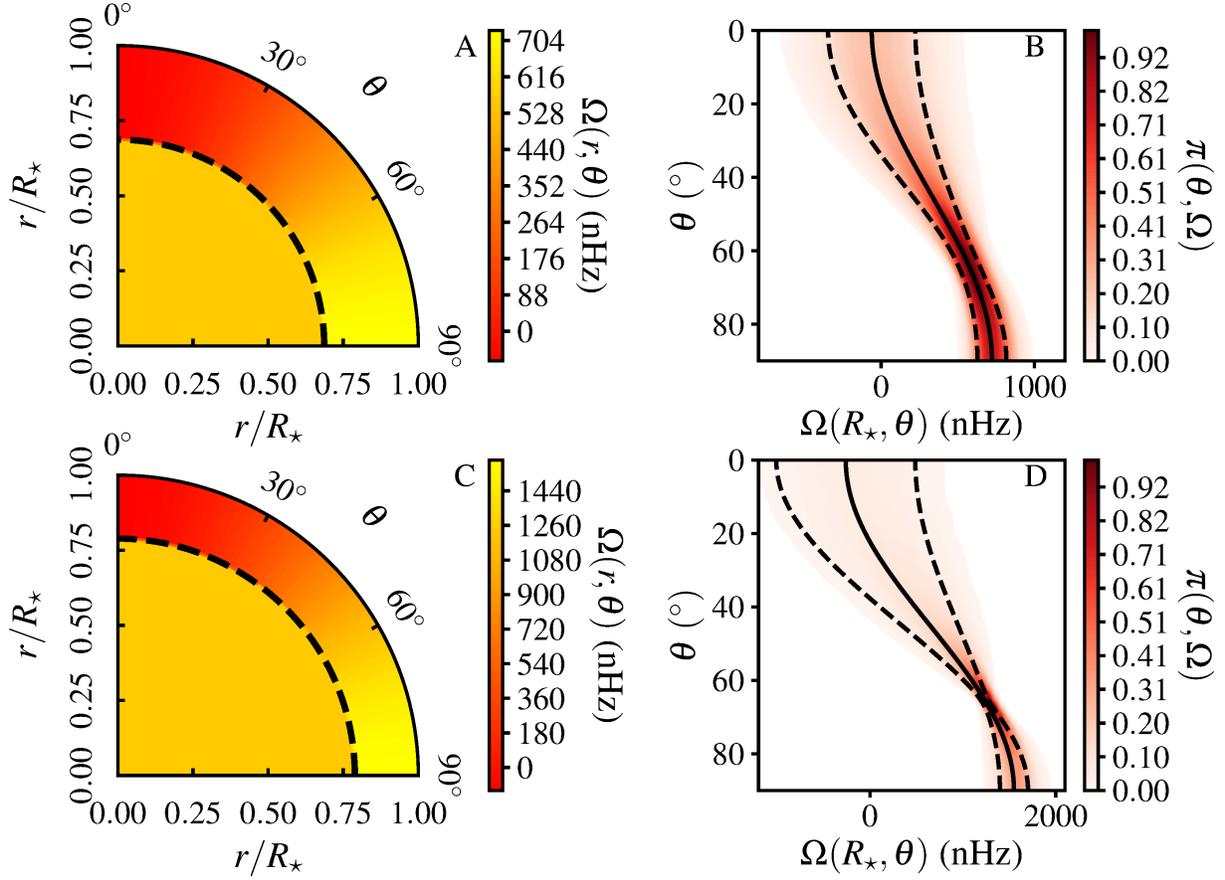, angle=-90, width=16cm}}
  \end{center}
\caption{\textbf{Rotation profiles from inversion for HD~173701 (A,B) and HD~187160 (C,D).}  Panels A,C display the most likely rotation profile (colors) and the interface between radiative and convective zone (dashed lines). Panels B,D show the probability density, $\pi(\theta,\Omega)$, at each latitude of the rotation profile in the convection zone (red-shaded region). The $1\sigma$ confidence interval is highlighted with dashed lines. The latitudinal differential rotation is well constrained for colatitudes $\theta>45^\circ$.}
\label{fig:rotation_profile}
\end{figure*}

Tables S2 and S3 contain the fundamental properties of the stellar models assumed for the inversion and the results of the inversion, respectively. Figure \ref{fig:rotation_profile} shows the two-dimensional rotation profiles and the probability density of the rotation rate at different latitudes for both stars. By construction, the profile is symmetric around the equator and around the rotation axis, so only one quadrant is shown. The rotation rate in the interior matches that of the envelope at a latitude of $26.5^{\circ}$ ($\theta=63.5^\circ$), due to the structure of the 
two-zone rotation profile. 
As is shown by the probability density of the rotation profile, the uncertainty on the rotation rate increases substantially toward the poles, beyond a value of approximately $45^{\circ}$ latitude. At high latitudes, the modes become much less sensitive to rotation, thus yielding less information. This limit is imposed by the lack of visible modes with angular degree higher than $l=2$ in the spectrum.  
We find that both HD~173701 and HD~187160 exhibit a latitudinal shear from equator to $45^{\circ}$ latitude $\Delta\Omega_{45}/\Omega_{\mathrm{eq}}$, that is approximately 5 times greater than that of the Sun.

As suggested by numerical simulations, the fast rotation of \arthur means that the functional form of its rotation profile may not be solar-like \cite{benomar2017science_supmat}.
We have considered this possibility for \arthur (Figure S8) and found that alternative rotation profiles have larger latitudinal shear than the solar-like rotation profile. However, computation of the Bayesian evidence (necessary to determine the goodness of the fit) shows that the solar-like profile is the most likely model.

\begin{figure*}[t]
  \begin{center}
	\subfigure{\epsfig{figure=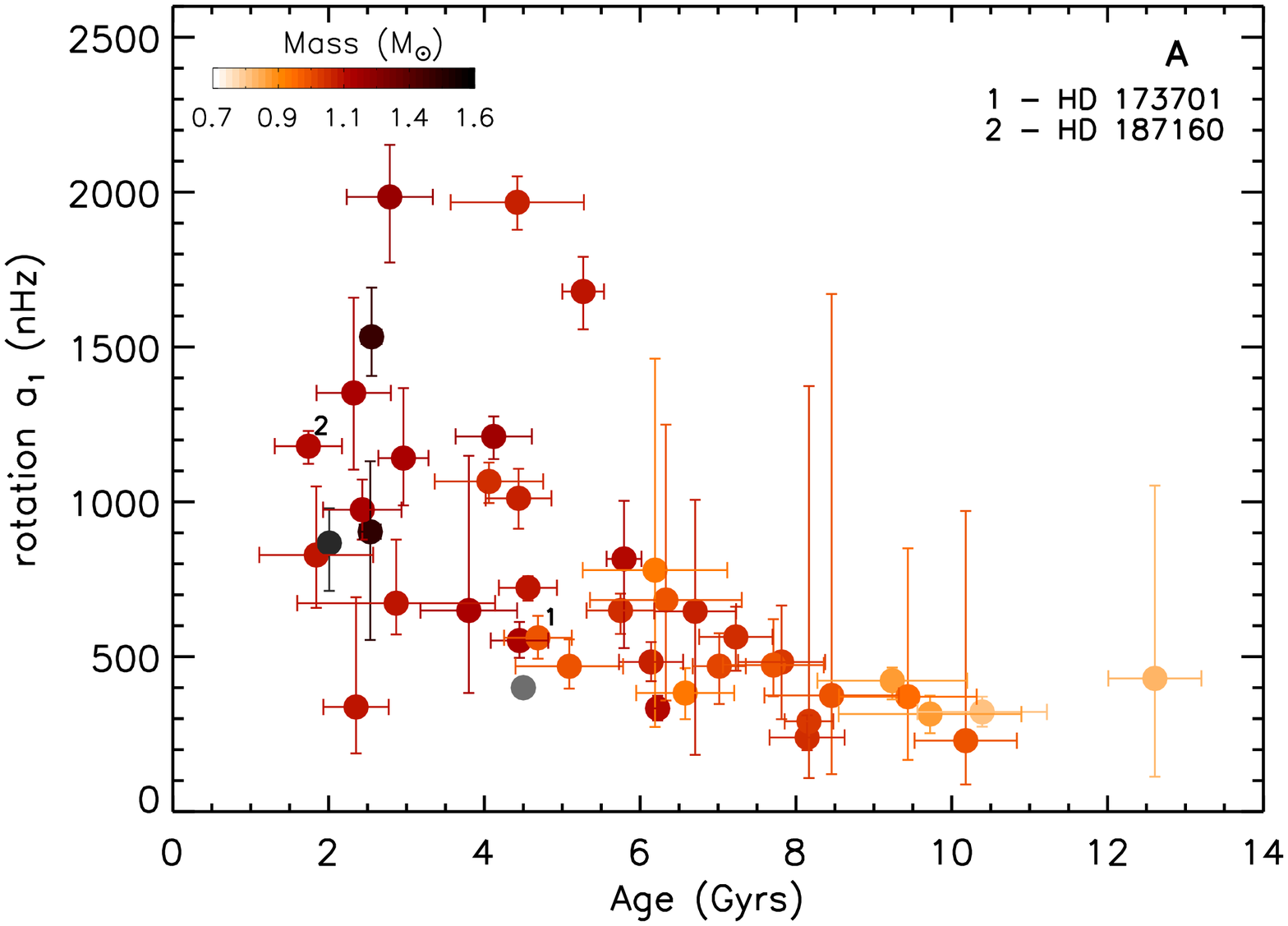, angle=0, width=7cm}}
	\subfigure{\epsfig{figure=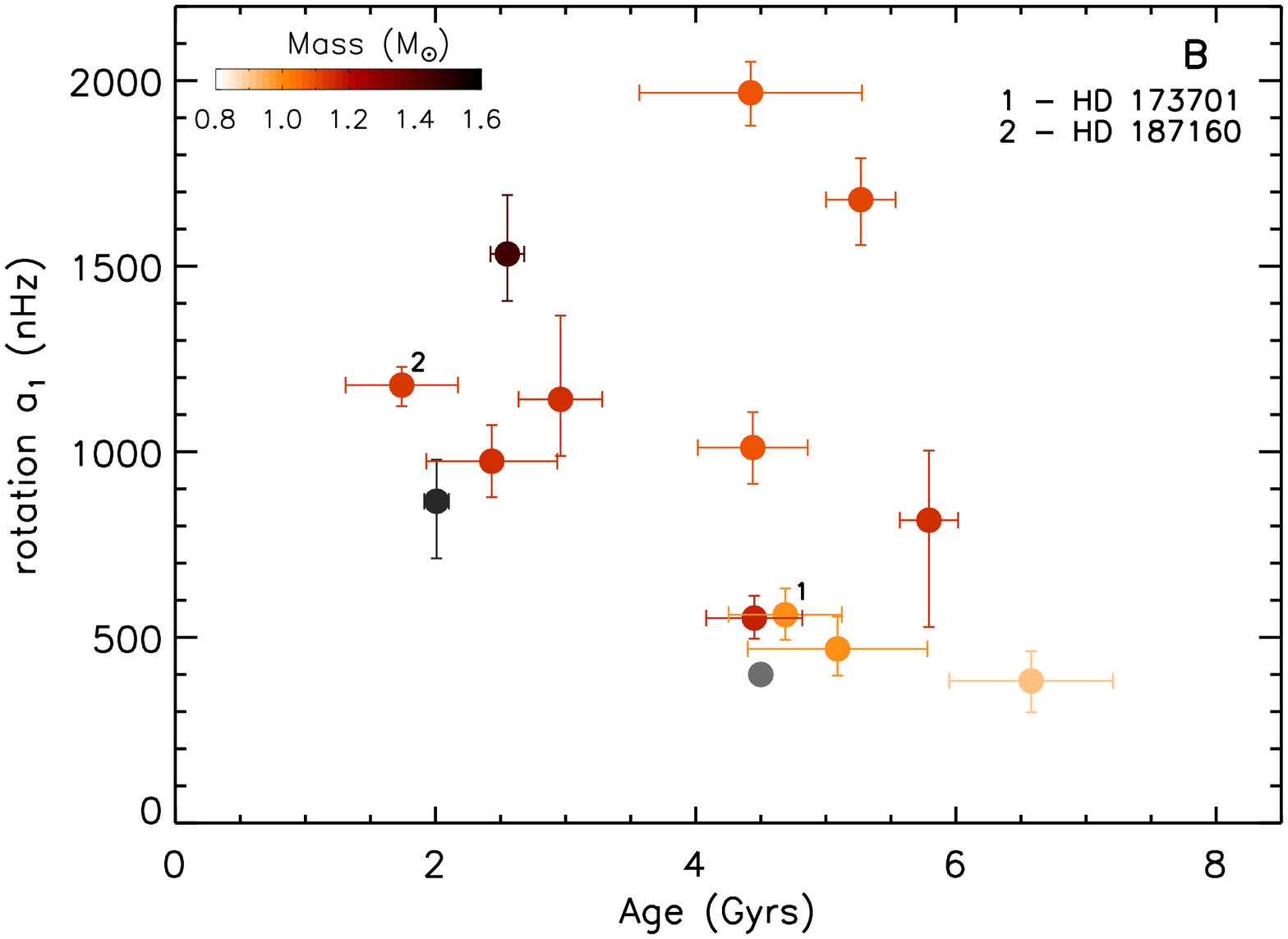, angle=0, width=7cm}} \\
     \subfigure{\epsfig{figure=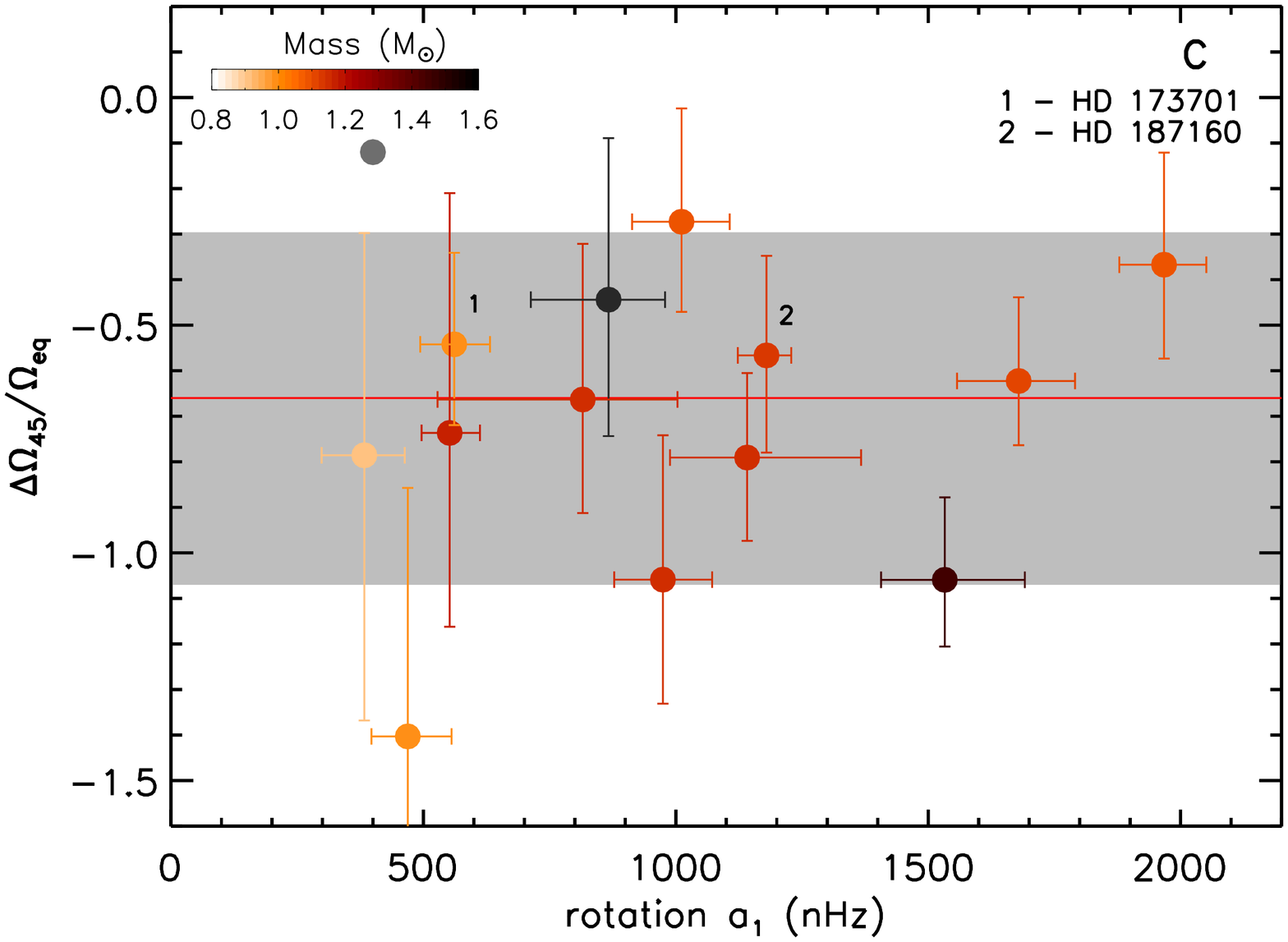, angle=0, width=7cm}}
\end{center}
\caption{\textbf{Internal rotation rate and differential rotation.} The solar {\it a}-coefficients are for the solar rotation profile (gray dot). \textbf{(A)} Measured values of $a_1$, as a function of age and mass for 40 stars. The well established decrease in rotation rate with stellar age (gyrochronology) is apparent.
\textbf{(B)} Same as (A) but for most significant detections.  \textbf{(C) The distribution} of $\Delta\Omega_{45}/\Omega_{\mathrm{eq}}$: the median is $64\%$. The gray region shows the $1\sigma$ dispersion.}
\label{fig:ensemble}
\end{figure*}

Due to loss of angular momentum, such as by magnetic braking, a star's rotation is expected to slow with its age \cite{Thompson2003}. 
This age-rotation relation is apparent in our data: Figure \ref{fig:ensemble}A,B show $a_1$ as a function of the age given by stellar models \cite{Aguirre2017}. 

As shown by Figure 3C, the latitudinal shear between the equator and mid-latitude for the subset of 13 stars with significant detections is $\approx 60\%$, albeit with a large scatter. HD~173701 and HD~187160 are representative of this ensemble, with a differential rotation of $\approx 50\%$. 
The Sun has a significantly lower shear factor than the considered ensemble. The difference is of more than $1\sigma$ from the dispersion of the ensemble.

This unexpectedly large shear poses a challenge to theoretical models. The balance between angular momentum transport (due to anisotropy in the turbulent flow) and small-scale flows, which act as enhanced turbulent viscosity, plays a dominant role in regulating latitudinal shear \cite{Hotta2016}. The  
large shear we find indicates a correspondingly large anisotropy in the turbulence, leading to efficient angular momentum transport and suppression of turbulent viscosity. In a typical stellar convection zone, turbulent anisotropy, driven by rotation, substantially affects large-scale flows. Thus, enhanced angular momentum transport and diminished turbulent viscosity amplify large-scale flows and suppress small-scale flows. This could be caused by a very efficient small-scale dynamo, in which small-scale flows are
suppressed \cite{Rempel2014}. 
 In addition, large-scale magnetic fields tend to reduce shear through the Lorentz force \cite{Karak2015}. However, our results indicate  that Lorentz-force feedback is ineffective in the  stars we investigated. Thus, the large-scale magnetic field is likely transported efficiently into the deeper regions of the star 
 in which rigid rotation is expected. For this, magnetic pumping is a candidate mechanism \cite{Brummell2002}.

\clearpage

\bibliographystyle{Science}

\textbf{Acknowledgments:} The authors thank T.R. White and S. Kamiaka for their helpful discussions. 

\textbf{Funding:} This work is supported by NYUAD Institute Grant G1502. Research funding from the German Aerospace Center (Grant 50OO1501) and the Max Planck Society (PLATO Science) is acknowledged.
Funding for the Stellar Astrophysics Centre is provided by The Danish National Research Foundation (Grant DNRF106). The research was supported by the ASTERISK project (ASTERoseismic Investigations with SONG and {\it Kepler}) funded by the European Research Council (Grant agreement no.: 267864). Funding for the {\it Kepler} mission is provided by the NASA Science Mission directorate. 

\textbf{Author contributions:} O.B. performed seismic analyses and lead in writing the manuscript. M.B. performed seismic inversion. M.B.N. analysed the lightcurves variability.  All authors discussed the results and contributed to sections of the manuscript.

\textbf{Competing interests:} None. 

\textbf{Data and materials availability:} This paper includes data collected by the {\it Kepler} mission provided by the Kepler Asteroseismic Science Operations Center (KASOC) and are in free access at \url{https://doi.org/10.7910/DVN/8SK6OL}. The analysis code is available at \url{https://github.com/OthmanB/TAMCMC-C}. Further details on the data are in Table S4.



\clearpage

\clearpage

\section*{Supplementary materials}

Materials and Methods

Fig. S1-S8

Table S1-S4

References (34-75)


\clearpage

\setcounter{table}{0} 
\setcounter{figure}{0} 
\setcounter{equation}{0} 
\renewcommand{\theequation}{S\arabic{equation}}
\renewcommand{\thefigure}{S\arabic{figure}}
\renewcommand{\thetable}{S\arabic{table}}

\begin{center}
	\section*{Supplementary Materials: Asteroseismic detection of latitudinal differential rotation in 13 Sun-like stars}
\end{center}

\section*{Materials and Methods} \label{text:method}

	\subsection*{Target selection and power spectrum preparation}

The ensemble of stars that are analysed in the present study are a subset of the LEGACY stars \cite{Lund2017}. Those correspond to stars with at least one year of continuous observation and with the highest signal-to-noise ratio among stars observed by the {\it Kepler} space instrument. From simulations using similar methods as in the section {\it Assessment of systematic errors} of this supplementary material, we found that these conditions are necessary but not sufficient to accurately and precisely measure the $a_3$ coefficient. The main critical condition is that the $l=0$ and $l=2$ modes must be well separated so that there is no crosstalk between modes. This excludes the hottest stars of the LEGACY samples for which the mode identification is problematic \cite{White2012,Benomar2009,Appourchaux2008}. The 40 analysed stars have effective temperatures approximately between 5250K and 6600K and metallicities between -0.7 and 0.4 \cite{Aguirre2017,Creevey2017}. However, there is an over-representation of stars with a higher temperature and lower metallicity than the Sun.

The lightcurve and power spectrum preparation was found to be important as several careful steps must be followed to reveal the faint stellar pulsations. Different procedures may lead to slightly different power spectra. Here, we proceeded with the analysis of the weighted (weight being the uncertainties on the flux) and un-weighted power spectra\cite{Handberg2014} provided by the KASOC pipeline (\url{kasoc.phys.au.dk}). These are prepared using pixel data with a specific mask adapted for asteroseismic {\it Kepler} targets and following the procedure of \cite{Handberg2014}. No additional modifications are made, such that these data are the same as those used in \cite{Lund2017,Aguirre2017} to derive the main fundamental parameters of the LEGACY stars. In addition, for HD~73701 and HD~187160, power spectra using the in-painting technique \cite{Pires2015} to mitigate the gaps in the timeseries were analysed. The results for $a_1$ and $a_3$ coefficient are found to be in agreement within $1\sigma$. Our results are computed using the un-weighted spectra.  

	\subsection*{Fitting method of the power spectrum}
	
The Maximum Likelihood Estimation approach (MLE) is often used to fit the power spectra \cite{Anderson1990}. However, it is mostly suitable for a likelihood function with a well-defined single maximum.
We use instead a Bayesian approach coupled with a Markov Chain Monte Carlo (MCMC) sampling algorithm. This method is known to be more adapted for measuring subtle effects in the data since it is less sensitive to converging on local maxima \cite{Benomar2009,Corsaro2014}. Furthermore, the MCMC method often provides more conservative and robust estimates on the uncertainties, as it returns the marginalized probability distribution functions of the parameters.

In a Bayesian approach the statistical criterion is the posterior density probability function $\pi(\boldsymbol{\theta} |  \boldsymbol{y}, \mathcal{M})$, built using conditional probabilities and Bayes' theorem,
\begin{equation} \label{eq:bayes}
	\pi(\boldsymbol{\theta}| \boldsymbol{y} , \mathcal{M}) \propto \pi(\boldsymbol{\theta} | \mathcal{M}) \pi(\boldsymbol{y}|\boldsymbol{\theta}, \mathcal{M}).
\end{equation}
Here, $\pi(\boldsymbol{y}|\boldsymbol{\theta}, \mathcal{M})$ is the probability of observing the data $\boldsymbol{y}$ given an underlying model $\mathcal{M}$, where the model is characterised by the parameters $\boldsymbol{\theta}$. Also, $\pi(\boldsymbol{\theta} | \mathcal{M})$ is the a priori knowledge of the fit parameters. 

The noise statistics for the power spectral density of independent bins $y_i$ at a given frequency $\nu_i$ is a $\chi^2$ distribution with two degrees of freedom (in fact, an exponential distribution), which gives the likelihood, 
\begin{equation} \label{eq:likelihood}
	\pi(\boldsymbol{y}|\boldsymbol{\theta}, \mathcal{M}) = \prod_{i=1}^N\frac{1}{\mathcal{M}(\nu_i, \boldsymbol{\theta})} \mathrm{e}^{- y_i/\mathcal{M}(\nu_i, \boldsymbol{\theta})},
\end{equation}
with $\mathcal{M}(\nu_i, \boldsymbol{\theta}) \equiv \mathcal{M}(\nu)$ being the model. Note that the physical model $\mathcal{M}$ that describes the discrete observations $\boldsymbol{y}$ is continuous. However, conditioned on $\mathcal{M}$, the observations are essentially independent. The assumption of independence is also justified by the very high duty cycle ($>95\%$) of the {\it Kepler} timeseries that we use. This means that the aliases and bin correlations in the power spectra due to the gaps in the time series are negligible. This was verified for CoRoT space-based observations \cite{Appourchaux2008}, a mission with a duty cycle of ~90\%. Furthermore, it is shown  that a degraded duty cycle enlarges the uncertainty on frequency parameters, which always remains higher than the Cram\'er-Rao bound (for the time series without gaps), but approaches this to within a fraction of a percent for a 95\% duty cycle \cite{Appourchaux1998,Stahn08}.
Therefore, the effects of the gaps on the power spectrum are negligible. 
The model is based on the sum of a noise background function $N(\nu)$, and of a sum of Lorentzian profiles \cite{Benomar2009},
\begin{equation}
\label{eq:power_Lorentzian}
\mathcal{M}(\nu) = \sum_{n,l,m} \frac{ H_{n,l,m} }{1 + (\frac{ \nu-\nu_{n,l,m}}{ 2 \Gamma_{n,l,m} })^2} + N(\nu).
\end{equation}
Here, $H_{n,l,m}$ and $\Gamma_{n,l,m}$, $\nu_{n,l,m}$ represent the height, width and frequency of the modes, respectively. 
As shown by Eq.~\ref{eq:freq:final}, $\nu_{n,l,m}$ is parametrized as a function of $a_1$, $a_3$ and $\beta_{n,l,m}$. In addition, the noise background is described using a Harvey-like profile and a white noise, as in\cite{Benomar2009,Appourchaux2008}. The mentioned quantities constitute the parameter space of dimension ranging from 50 to 100, depending on the number of measured pulsation modes. Although the parameter space is apparently large, note that the observations $\boldsymbol{y}$ contain typically $10^5$ data points, which greatly exceed the number of parameters. Furthermore, the use of priors limits the sampled volume.

\subsection*{MCMC sampling}
The analysis of the power spectrum was made using an MCMC algorithm, based on a Metropolis-Hasting scheme, with parallel tempering \cite{Metropolis1953,Hastings1970,Earl2005}. The number of parallel chains is fixed to $c_{max}=10$, with a temperature distribution following a geometric law $T=\lambda^{c}$, fixing $\lambda$ such that the maximum temperature is $T_{max}=100$ (and $T_{min}=1$).  A total of 2 million samples are acquired (for each chain), after a burn-in phase of 100 000 samples and a training phase of 700 000 samples. 
This ensures that we reach an acceptance rate near 0.234 that remains stable along the acquisition phase (within $\simeq 10\%$). The training phase uses an adaptive algorithm based on \cite{Atchade2006} to optimize the covariance matrix of the (Gaussian) proposal probability density function. 
More details are in \cite{Benomar2009} and the implementation in C++ is available at \url{https://github.com/OthmanB/TAMCMC-C}.
We verify that the target distribution is reached by using different diagnostics. First, we compare results from the first half of the samples with the full set. We did not see changes that exceed $~1\%$ percent in the statistical indicators (median and confidence interval). A visual inspection of the probability density functions is also performed.
Secondly, we use the Heidelberger and Welch test \cite{Heidelberger1983,Heidelberger1981} and the Geweke test \cite{Geweke92}. These are quantitative convergence tests that are applied to each of the parameters. We used their implementation in R from the \textsc{CODA} library \cite{CODA2006}, and the \textsc{py-coda} python module (\url{https://github.com/surhudm/py-coda}). The default parameters were used. We found that $\sim 500\,000$ samples are required to pass both convergence tests for all parameters of our spectral model. This is below the number of samples we perform for each star, ensuring that we are properly sampling the target distributions.

\subsection*{The effect of rotation on pulsation frequencies}

The rotation of the star lifts the frequency degeneracy of non-radial modes ($m$, $l\neq 0$), 
producing a multiplet of $2l+1$ modes of azimuthal orders $m$.

For slow and moderate rotators (rotation periods larger than a few days), the rotation is a small perturbation $S_{n,l,m}$ relative to the spherically symmetric pulsations frequencies $\nu_{n,l}$,
\begin{equation}
\nu_{n,l,m} = \nu_{n,l}  + \,m\, S_{n,l,m} + O(\Omega^2).
	\label{eq:rot:generality}
\end{equation}
The perturbation $S_{n,l,m}=(\nu_{nlm} - \nu_{nl-m})/2m$ is the symmetric rotational splitting \cite{Goupil2004} and $O(\Omega^2)$ represents higher-order effects of the rotation (e.g. asphericity). 

To first order, the perturbation on the frequency due to rotation is
\begin{equation}
	S_{n,l,m} = \frac{1}{2 \pi} \int^{R_\star}_0 \int^\pi_0 K_{n,l,m}(r,\theta) \, \Omega(r,\theta) r dr d\theta.
		\label{eq:rot:order1}
\end{equation}
Here, ${R_\star}$ is the radius of the star and the kernel $K_{n,l,m}(r,\theta)$ \cite{Schou1994,Aerts2010} determines the sensitivity of the mode $(n,l,m)$ to the rotation at the radial point $r$ and co-latitude $\theta$. 

Here we also consider the second-order perturbation which accounts for the asphericity of the star. The centrifugal force distorts the oscillation mode cavities, inducing an additional frequency perturbation. This term scales as $ \Omega^2(r, \theta=\pi/2) / (\mathcal{G} \rho_\star)$, where $\mathcal{G}$ is the gravitational constant and $\rho_\star$ is the mean density of the star.

Other more complicated perturbations such as a large-scale magnetic field \cite{Dziembowski1991} may also modify the shape of the mode cavities. 
Naming these various perturbation terms $\epsilon(\nu_{n,l})$, the second-order effects of rotation on the mode frequencies can be expressed as,
\begin{align}
	\beta_{n,l,m}(\nu_{n,l}) & = \frac{4 \pi \, Q_{lm} } {3 \, \mathcal{G} \rho_\star} \, \Delta_{n,l} \, \nu_{n,l} \, \Omega^2(r, \theta=\pi/2)  + \epsilon(\nu_{n,l}) \nonumber \\
					  & = \eta_{n,l,m}  + \epsilon(\nu_{n,l}).
	\label{eq:rot:order2}
\end{align}
with $Q_{l,m} = \frac{l(l+1) - 3m^2}{(2l-1)(2l+3)}$ \cite{Kjeldsen1998} and $\Delta_{n,l} \approx 3/4$ \cite{JCD2000,Aerts2010}.
The term $\beta_{n,l,m}(\nu_{n,l})$ should be added to Eq.~\ref{eq:rot:generality} to have the total perturbation due to rotation and to the asphericity of the star.

When modelling the pulsation frequencies, we investigate cases with only the centrifugal term $\eta_{n,l,m}$ and compare with cases when $\epsilon(\nu_{n,l}) \neq 0$. This enables us to evaluate whether other forces than the centrifugal force distort the star. 

In this study, the final objective is to solve an inverse problem, which consists of inferring the rotation profile based on measures of $S_{n,l,m}$.  In the early days of helioseismology,  the Clebsch-Gordon \emph{a}-coefficient decomposition \cite{Duvall1986, Brown1989, Ritzwoller1991, Schou1994} was used to describe the splitting $S_{n,l,m}$ in terms of a sum of polynomial coefficients $a_{1}, a_{2}, ..., a_{k}$. The decomposition associates the odd coefficients with the characteristics of the rotation, such as the differential rotation.

In the \textit{Kepler} observations of \arthur and HD~173701, only $l \le 2$ modes could be precisely measured. This limits the number of measurable $a$-coefficients to $a_{1}$ and $a_{3}$, which we subsequently use in our inversion for the rotation profile $\Omega(r,\theta)$.

The symmetric splitting for $l=1$ and $l=2$ modes is a combination of \emph{a}-coefficients \cite{Gizon2004},
\begin{align}
	S_{n,1,1} &= a_1(n,1), \\
	S_{n, 2, m} &= a_1(n, 2) + \frac{1}{3} ( 5 m^2 - 17) a_3(n, 2).
\end{align}
In order to improve the robustness of the estimates, we simplify these equations by assuming $S_{n , 1, 1} = S_{n, 2, 2}$, which leads to $a_1(n, 1) = a_1(n, 2) +  a_3(n,2)$. Looking at models of the stars, we could verify that $S_{n , 1, 1}$ differs by only $\approx 5\%$ from $S_{n , 2, 2}$.  Finally, noting that the dependence on $n$ of $a_1$ (see figure \ref{fig:a1n}) is in practice difficult to measure \cite{Nielsen2014},
the perturbed frequencies $\nu_{n,l,m}$ are now approximated by,
\begin{equation}
	\nu_{n,l,m} \simeq \nu_{n,l} + m a_1 + \beta_{n,l,m}(\nu_{n,l}) + C_{l,m} a_3,
	\label{eq:freq:final}
\end{equation}
with $a_1 = \overline{a_1(n, 2)}$, $a_3=\overline{a_3(n ,2)}$ are averages over $n$; with $C_{1,m} = m$ and $C_{2,m} = \frac{5m^3 - 17m}{3}$. 
Eq.~\ref{eq:freq:final} is the relation that is used  when fitting the power spectrum with the method described in the section {\it Fitting method of the power spectrum}. Note that the derivation of the rotation profile using the observed \emph{a}-coefficients requires another step, described in the section {\it Seismic inversion for the rotation profile from $a_1$ and $a_3$}.

The choice of the prior $\pi(\boldsymbol{\theta} | \mathcal{M})$ is important in a Bayesian framework. Specifics concerning the choice of priors in the case of a solid-body rotation are already described in several studies \cite{Benomar2009,White2017}. However, our model differs in terms of rotation as it involves the measure of the average internal rotation  $a_1$, the asphericity $\beta_{n,l,m}(\nu_{n,l})$ and the latitudinal differential rotation by means of  $a_3$. Priors on each of these parameters are justified below.

\subsection*{Prior on $a_1$} \label{sec:priors:a1}

We choose a non-informative uniform prior on $a_{1}$ of the form,
\begin{align}
	p( a_1 | \mathcal{M}) = 
	\begin{cases}
		\frac{1}{ a_1^{\rm max}}  & \mbox{ if } 0 < a_1 < a_1^{\rm max} \\
		0 & \mbox{ otherwise.}
	\end{cases}
\end{align}
The average internal rotation rate is expected to be below a few days so that $a_1^{\rm max}$ is fixed to $5\,\mu$Hz (corresponding to $2.3$ days).

	\subsection*{Prior on $\beta_{n,l,m}(\nu_{n,l})$ when $\epsilon(\nu_{n,l})=0$} \label{sec:priors:beta:1}
	
Here, we consider the special case for $\epsilon(\nu_{n,l})=0$. Then the only perturbation to the frequency is $\eta_{n,l,m}=\frac{4\pi \, Q_{l,m}} {3\,\mathcal{G} \rho_\star} \, \Delta_{n,l} \, \nu_{n, l} \,  \Omega^2(r, \theta=\pi/2)$, and so it is possible to approximate $a_1$ as $\Omega(r, \theta=\pi/2)$. 
Modes of high adjacent radial order and of the same degree $l$ are separated by an almost uniform frequency spacing $\Delta\nu$, the so-called large separation. 
 The large separation is related to the sound speed $c(r)$ inside the star and is sensitive to the mean stellar density $\rho_\star$. Assuming that the solar structure scales with other Sun-like stars, one can relate the mean solar density $\rho_\odot=(1.4060 \pm 0.0005)\times 10^3$\,kg\,m$^{-3}$ and its mean frequency spacing $\Delta\nu_\odot = 135.20 \pm 0.25$ $\mu$Hz \cite{Garcia2011b}, with those of other stars,
\begin{equation}
	\frac{\rho_{\star,\mathrm{s}}}{\rho_\odot } = \frac{\Delta\nu^2}{\Delta\nu^2_\odot}. 
	\label{eq:scaling:dnu}
\end{equation}

This allows us to re-formulate $\eta_{n,l,m}$ as,
\begin{equation}
	\eta_{n,l,m} = \frac{4 \pi \, \Delta\nu^2_\odot} {3 \, \mathcal{G} \, \rho_\odot \, \Delta\nu^2} \, Q_{l,m} \, \Delta_{n,l} \, \nu_{n, l} \,  a^2_1 = \eta_0 \, Q_{l,m} \, \Delta_{n,l} \, \nu_{n,l} \, a^2_1,
\end{equation}
defining $\eta_0=\frac{4 \pi \, \Delta\nu^2_\odot} {3 \, \mathcal{G} \, \rho_\odot \, \Delta\nu^2}$.

The relative difference in radius between the equator and the pole $\Delta R/R$ of the star is then simply given by,
\begin{equation}
	\frac{\Delta R}{R} = \frac{3}{8 \pi} \eta_0 a^2_1.
\end{equation}

Noting that $\Delta\nu$ is known a priori with a typical error of only $\approx 0.5 \%$, while $a_1$ is constrained to $\approx 5 \%$, it is clear that the main source of uncertainty is $a_1$. We therefore fix $\eta_0$, which is equivalent to the prior,
\begin{equation}
	p(\eta_{n,l,m} | \mathcal{M}) = \delta( \eta_{n,l,m} - \eta_0) \, Q_{l,m} \, \Delta_{n,l} \, \nu_{n,l} \, a^2_1,
\end{equation}
where $\delta$ is the Dirac delta function. 

	\subsection*{Prior on $\beta_{n,l,m}(\nu_{n,l})$ when $\epsilon(\nu_{n,l}) \neq 0$} \label{sec:priors:beta:2}

Several mechanisms other than the centrifugal force may affect the asphericity of the mode cavity, and so models with $\epsilon \neq 0$ are also considered in this study. Noting that $\beta_{n,l,m}(\nu_{n,l})$ contains all of the information about the asphericity, the most general analysis strategy consists in fitting directly for $\beta_{n,l,m}(\nu_{n,l})$. Disentangling $\epsilon(\nu_{n,l})$ from the centrifugal term can then be done \emph{a posteriori}. One must first decide upon a functional form for $\beta_{n,l,m}(\nu_{n,l})$; here we use 
\begin{equation}  \label{eq:asph}
	\beta_{n,l,m}(\nu_{n,l}) = \beta_0 Q_{lm} \nu_{n,l},
\end{equation}
where  $\beta_0$ is a free parameter.
With this expression, $\beta_0$ can be directly compared to the coefficient $\eta_0 a_1^2$ of the centrifugal force, so that the presence of additional forces can be easily determined.

An examination of Eq.~\ref{eq:asph} shows that an asphericity mostly shifts the $m=0$ component either to higher frequency (for an oblate star) or to lower frequency (for a prolate star), compared to the frequencies of a spherically symmetric star. 
Considering Eq.~\ref{eq:freq:final} lets us define a limit on $\beta_0$,
\begin{equation}
	|\beta_0| \leq C \frac{a_1+ a_3}{(Q_{11} - Q_{10}) \nu_{n,1}},
\end{equation}
with $0<C<1$ as an adjustable parameter. When $C=1$, $\nu_{n,1,m=0} = \nu_{n,1, m=1}$. 
This possibility is rather extreme and not expected in slow rotators, for which the asphericity caused by the centrifugal force is $\Delta R / R \approx 10^{-5}$. At solar density and with pulsation frequencies comparable to the Sun, this represents at most a frequency shift of the zonal component $\nu_{n,l, m=0}$ of 100 nHz. This has to be compared to a rotational splitting of $\approx 2000$ nHz for a solar-like star rotating in a few days. In our analysis we consider instead $C=0.5$.  
It is evident that the maximum of $a_1$ ($a_1^{\rm max}$) and the minimum of $\nu_{n,1}$ ($\nu_{n,1}^{\rm min}$) must be known a priori in order to use this condition. These are obtained by fitting the power spectrum of the star with fixed $\epsilon(\nu_{n,l})=0$ and $a_3 =0$. 
Note that $a_3 \ll a_1$ such that the maximum value of $a_3$, $a^{\rm max}_3 \approx 0$ is assumed hereafter.
Finally, setting an uniform prior on $\beta_0$ gives
\begin{align}
	p( |\beta_0| | \mathcal{M}) = 
	\begin{cases}
		\frac{1}{ \beta^{max}_0}  & \mbox{ if } 0 < |\beta_0| < \beta^{max}_0 \\
		0 & \mbox{ otherwise,}
	\end{cases}
\end{align}
with $\beta^{max}_0 = C \frac{a^{max}_1}{(Q_{11}-Q_{10}) \nu^{min}_{n,1}}$.
 
	\subsection*{Prior on $a_3$} \label{sec:priors:a3}
	
Because the amplitude of the latitudinal differential rotation is expected to represent only a fraction of the average rotation rate, one can consider $a_3 \ll a_1$. Furthermore, the Sun has evidently a faster rotation at the equator than at the pole (which corresponds generally to $a_3 > 0$), but the opposite scenario (case with $a_3<0$) cannot be excluded a priori in other stars \cite{Gastine2014}. The chosen prior must therefore consider both positive and negative solutions of $a_3$. Examination of Eq.~\ref{eq:freq:final} shows that $a_3$ is a location parameter \cite{Jeffreys1961}. Location parameters have probability density function invariant by translation. Thus, an adequate non-informative prior is a uniform prior. Due to all these considerations, we impose
\begin{align}
	p( a_3 | \mathcal{M}) = 
	\begin{cases}
		\frac{1}{ 2 \, a^{max}_{3}}  & \mbox{ if } - a^{max}_{3} < a_3 < a^{max}_{3} \\
		0 & \mbox{ otherwise.}
	\end{cases}
\end{align}
For the Sun, $a_3/a_1 \simeq 1.2\%$, caused by a latitudinal differential rotation between the equator and $45^\circ$ of latitude of $\simeq 11\%$. Eq.~\ref{eq:a3a1} shows that the importance of the latitudinal differential rotation is actually proportional to $a_3/a_1$. Therefore, $a_3/a_1$ is expected to be of the same order in other solar-like stars.
Due to this, the value $a^{max}_{3}=220$ nHz is chosen for stars with $a_1>800$ nHz. For slower rotators, $a^{max}_{3}=150$ nHz.  

	\subsection*{Seismic inversion for the rotation profile from $a_1$ and $a_3$}
    
    Once \emph{a}-coefficients are measured from the power spectrum using Eq.~\ref{eq:bayes}-\ref{eq:power_Lorentzian} (see the sections {\it Fitting method of the power spectrum} and {\it The effect of rotation on pulsation frequencies}, for details), one needs to translate them into a stellar rotation profile. This section explains this procedure.

Choosing orthogonal polynomials implies that the \emph{a}-coefficients correspond one-to-one to a projection of $\Omega(r, \theta)$ onto polynomials $\psi_i(\theta)$,
\begin{equation}
	\Omega(r, \theta) = \sum^{s_{max}}_{s=0} \Omega_s(r) \psi_s(\theta),
		\label{eq:rot:omega:expanded}
\end{equation}  
with $\Omega_s(r)$ is a radial rotation profile, to be chosen \cite{Ritzwoller1991, Schou1994, Pijpers1997}. 

Only two measurements are available to us ($a_1$ and $a_3$) so in order to resolve the inverse problem without facing degeneracies of parameters, we are limited to a two-zone model such that $\Omega(r, \theta) = \Omega_0(r) \psi_0(\theta) + \Omega_1(r) \psi_1(\theta)$ where \cite{Gizon2004}
\begin{align}
	\psi_0(\theta) & = 1, \\
	\psi_1(\theta) & = \frac{3}{2} ( 5 \cos^2 \theta - 1).
\end{align}
These choices of $\psi_0$ and $\psi_1$ ensure that there exists a one-to-one relation between the $a$-coefficients and the rotation parameters $\Omega_0(r)$ and $\Omega_1(r)$. 
Since the Sun has nearly-uniform rotation in the radial direction \cite{Schou1998, Thompson2003, Darwich2013}, and a latitudinal rotation gradient in the convective zone, an appropriate choice is to define the zone boundary at the interface between the radiative and convective zone. By using this model we can write,
\begin{align} \label{eq:domega0}
	\Omega_0(r) & = \otz \mbox{ and,} \\  
	\Omega_1(r) & =
	\begin{cases}
		\oto  & \mbox{if } r_c \leq r \leq R_\star \\
		0  & 0 \leq r \leq r_c \mbox{ otherwise,}
	\end{cases} \label{eq:domega1}
\end{align} 
with $\otz$ the average rotation rate of the star and $\oto \, \psi_1(\theta)$ the latitudinal rotation rate. 
This profile assumes that the radiative zone does not rotate notably faster than the convective zone. This is consistent with most recent observations of main-sequence stars\cite{Kurtz2014,Saio2015,Benomar2015,Nielsen2017}. In the convective zone, the rotation profile can be recast as a function of the contrast of rotation between the equator and pole $\Delta\Omega=\Omega_{\rm pole} - \Omega_{\rm eq} = 15 \oto /2$ and the equatorial rotation, \begin{equation}
	\Omega = \Omega_{\rm eq} + \Delta\Omega \cos^2 \theta.
\end{equation}
Here $\Omega_{\rm eq}$ is the equatorial rotation and $\Omega_{\rm pole}$ is the polar rotation.
By using the definition of the rotational splitting,
\begin{equation}
	2 \pi\,S_{n,l,m}= \int_0^{R_{\star}} \int_0^\pi K_{n,l,m}(r, \theta) \, \Omega_i(r) \, \psi_i(\theta) \, r \, dr d\theta,
		\label{eq:kernel:final}
\end{equation}
and noting that $a_3 = (S_{n,2,2}  - S_{n,2,1})/5$, we obtain that 
\begin{equation}
	2 \pi\,a_1 = \otz \int^\pi_0 \int^{R_\star}_{0} K_{2,2}(r, \theta) r dr d\theta,
		\label{eq:a1:final}
\end{equation}
 and
\begin{equation}
	2 \pi\,a_3 = \frac{\oto}{5} \int^\pi_0 \int^{R_\star}_{r_c} (K_{2,2}(r, \theta) - K_{2,1}(r, \theta)) \psi_1(\theta) r dr d\theta,
		\label{eq:a3:final}
\end{equation}
where $K_{2,2}(r, \theta)$ and $K_{2,1}(r, \theta)$ are kernels averaged over the range of observed radial orders.

Since the integrals in Eq.~\ref{eq:a1:final} is $\approx 1$ \cite{Aerts2010}, we obtain
\begin{equation}
	\frac{a_3}{a_1} \simeq \frac{\oto}{5 \Omega_0} \int^\pi_0 \int^{R_\star}_{r_c} (K_{2,2}(r, \theta) - K_{2,1}(r, \theta)) \psi_1(\theta) r dr d\theta \propto \frac{\Delta \Omega}{\Omega_0}.  \label{eq:a3a1}
\end{equation}

Eqs.~\ref{eq:a1:final} and \ref{eq:a3:final} can be inverted once the kernels are known. 
This will give us the probability densities of the coefficients $\otz$ and $\oto$. 
To do this we first need to evaluate the integrals of the rotation sensitivity kernels $K_{2,2}$ and $K_{2,1}$ to compute the mass density profile of the stars and their oscillation mode eigenfunctions and determine $K_{2,2}$ and $K_{2,1}$. This can be achieved by using a stellar evolution and oscillation codes to find the best stellar model that describes seismic (frequencies) and non-seismic observables (log(g), effective temperature, etc.).  Such a computation was done in \cite{Aguirre2017} for the LEGACY project, a project that focuses in modeling stars observed by \textit{Kepler} for longer than a year. 
Here, we use stellar models from \cite{Aguirre2017} based on the stellar evolution code \textsc{astec} \cite{JCD82b,JCD08a} and the pulsation code \textsc{adipls} \cite{JCD08b}. The optimisation procedure used to find the best stellar models is named \textsc{astfit} by their authors.

	\subsection*{Testing cylindrical rotation profiles for HD~187160}
    
   For fast rotating sun-like stars, axisymmetric mean-field simulations and some 3D convection simulations \cite{Gastine2013} suggest that the morphology of the rotation profile may differ from the Sun. These simulations, which parametrize the angular momentum transport through anisotropic turbulent Reynolds stresses, tend to have rotation that is almost constant on cylinders for very fast rotation rates. Then the rotation depend largely on the distance to the rotation axis. The differential rotation profile in the convection zone then depends on the distance $\varpi$ to the rotation axis, with $\varpi = r \sin(\theta)$. 
    For example, some studies \cite{Kuker2011} found that for a rotation period of 4.5 days (corresponding to 6 times faster than the solar rotation rate $\Omega_\odot$) the rotation profile shows both solar-like and cylindrical characteristics in the convective zone. It becomes almost completely cylindrical for $\Omega \approx 10 \Omega_\odot$. On the other hand, it has been suggested that cylindrical profiles may appear for rotation rates as low as $\Omega \approx 2 \Omega_\odot$ \cite{Hotta2011}, with a weak gradient of rotation for $\varpi > r_c$ and a rotation that increases outward almost linearly for $\varpi < r_c$. In these simulations, the rotation rate is almost constant inside the cylinder. From the simulations, $r_{\rm c}$ is typically the position of the transition between the convective outer layers and the inner radiative layers, called the tachocline. 
    Other simulations found a cylindrical rotation in the convective zone when $\Omega \geq 3 \Omega_\odot$ \cite{Brown2008}, but with a rotation that varies with $\varpi$ throughout the entire convection zone. 
    
    The latitudinal entropy gradient in the bulk of the convection zone plays an important role determining the details of the rotation profile \cite{Rempel2005}. As the rotation rate increases the effect of the latitudinal entropy gradient becomes less important and rotation will tend to be constant on cylinders everywhere. In particular, the rotation rate inside the 
cylinder in the convection zone will be that of the radiative interior (which is usually assumed to be constant).
 
Despite these differences, most models suggest that rotation should be faster at the equator than at higher latitudes for stars that rotate faster than the Sun \cite{Brun2017}. 
Asteroseismology may provide a means of verifying the plausibility of such simulations. 
The cylindrical profiles suggested by the mean-field and convection simulations can be written as
\begin{align}
	\Omega(r, \theta) & = 
	\begin{cases}
		& \Omega_{\rm rad}  \mbox{  if } 0 \leq r \leq r_c \\
	    & \Omega_{\rm CZ}(\varpi)  \mbox{  otherwise.}
	\end{cases} \label{eq:domega_cyl0}
\end{align}
In this equation $\Omega_{\rm rad}$ is the rotation rate inside the radiative zone and $\Omega_{\rm CZ}(\varpi)$ is the rotation profile within the convective zone.
The angular velocity may vary linearly with distance from the rotation axis throughout the convection zone such that
\begin{equation}  \label{eq:domega_cyl1}
	\Omega_{CZ}(\varpi) = \Omega_{\mathrm{eq}} - ( 1 - \varpi/R)(\Omega_{\mathrm{eq}} - \Omega_{\mathrm{pole}})   \mbox{ for } r \geq r_c.
\end{equation}
Here, $\Omega_{\mathrm{eq}}$ and $\Omega_{\mathrm{pole}}$ denote the equatorial and polar rotation rate.

Alternatively the rotation can also be constant inside the 
cylinder, meaning that there is no change in the rotation profile when crossing the transition between the convective zone and the radiative zone,
\begin{align}
	\Omega_{CZ}(\varpi) & = 
	\begin{cases}
		& \Omega_{\mathrm{eq}} - \frac{1 - \varpi/R}{1 - r_c/R} (\Omega_{\mathrm{eq}} - \Omega_{\rm rad})  \mbox{  if } r \geq r_c \mbox{ and } \varpi \geq r_c \\
	    & \Omega_{\rm rad}   \mbox{  if } r \leq r_c \mbox{ and } \varpi \leq r_c \\
	\end{cases} \label{eq:domega_cyl2}
\end{align}
This leads to the absence of a tachocline such that unlike the Sun, the radiative interior is not rotating uniformly.

The model described by Eq.~\ref{eq:domega_cyl1}, which maintains the presence of a tachocline is perhaps better adapted to the transition between slow and faster rotation, while the second case (Eq.~\ref{eq:domega_cyl2}) is for even faster rotation.

The parameters in Eq.~\ref{eq:domega_cyl1} ($\Omega_{\rm rad}$, $\Omega_{\mathrm{eq}}$, and $\Omega_{\mathrm{pole}}$) and Eq.~\ref{eq:domega_cyl2} ($\Omega_{\rm rad}$, $\Omega_{\mathrm{eq}}$) are estimated in a Bayesian framework for HD~187160.
The likelihood $f(\Omega(r,\theta)| a_1, a_3)$ is chosen as a bivariate normal density. The moment of the likelihood probability density are estimated from the MCMC samples obtained from the fit of the power spectrum.

We attempted to compute the posterior distributions for the most realistic model that accounts for a tachocline (Eq.~\ref{eq:domega_cyl1}). However, since there are only two measurements ($a_1$ and $a_3$), we observe significant degeneracies between the three parameters of the model. It is therefore not possible to find unique solutions for the model parameters. 
Measurements of e.g. $a_5$ would likely reduce the degeneracies between parameters and ensure the uniqueness of the solution. However this would only be possible in the exceptional cases where $l=3$ mode are visible and of high amplitudes.

The posterior distributions of the model parameters with tangent cylinders and no tachocline is shown in Figure \ref{fig:rot:cyl1}. 
Here, the solution is well defined with $\Omega_{\mathrm{eq}} = 2452 \pm 603$ nHz and $\Omega_{\rm rad} = 949 \pm 197$ nHz. 
We determine that the shear between $45^\circ$ and the equator is $\Delta\Omega_{45}/\Omega_{\mathrm{eq}} \approx 1.00$.

This is a stronger differential rotation than the solar-like model. To evaluate which model could be the most compatible with the observations, the Bayesian evidence is computed. We find that the solar-like model is 1.7 times more likely. This is a substantial difference but not strong enough to decisively reject a rotation in cylinders \cite{Jeffreys1961}.

	\subsection*{Assessment of systematic errors} \label{appendix:2}
	
An earlier study \cite{Gizon03} evaluated the possibility of measuring $a_3$. However, they did not consider a global fit. They investigated the reliability of $a_3$ by fitting a frequency range $\Delta\nu$, with a single set of $l=0, 1, 2$ modes, and assuming an average rotation rate two to six times that of the Sun. They also considered solar mode lifetimes such that for faster rotation rates, the split components become well separated. Under these conditions, measurements of $a_1$, $\beta_{n,l,m}(\nu_{n,l})$ and $a_3$ are unbiased and reliable within a range of inclination $35 < i < 75$ degrees. In light of the latest {\it Kepler} observations \cite{Lund2017}, it is now evident that these conditions are optimistic. The main reason is that fast rotation is often associated with hot stars \cite{Benomar2015}, which typically have short-lived pulsations. Thus, mode blending may prevent us from separating the split components and achieving the required level of precision to determine $a_3$. Furthermore, the methodology for analysing solar-like pulsations involves the simultaneous fit of all statistically significant modes of the power spectrum, contrary to \cite{Gizon03}.
	 
It is therefore important to assess the potential systematics on rotation parameters inferred from seismology with realistic simulations. This is achieved by generating artificial spectra that use the same signal-to-noise ratio, mode widths ($\Gamma_{n,l}$) and frequencies ($\nu_{n,l}$) as those obtained when fitting the power spectra with $a_3=0$ (see equation \ref{eq:power_Lorentzian}). The frequency resolution is also fixed to that of the observations.

Contrary to the real data, the systematic error on $a_1$ and $a_3$ is obtained by analysing the artificial noise-free power spectrum (often referred as to the limit spectrum).  A grid of spectra is constructed spanning the range $a_3=[-150, 150]$~nHz, $a_1=[1100, 1300]$~nHz and $i=[0,90]$ degree for HD~187160. A similar grid is constructed for HD~173701 but with $a_1=[350, 750]$~nHz. Each of the spectra is fit using the same method and algorithm as for real observations. This gives the probability density function (PDF) for each parameters in the fit, including $a_1$, $a_3$ and $i$. The biases are defined as 
\begin{equation}
	b(x^{(\rm true)}) = x^{\rm (obs)} - x^{\rm (true)},
\end{equation}
where $b(x^{(\rm true)})$ is the systematic error for the parameter $x$ (either $a_1$ or $a_3$) estimated from the grid. It is estimated using the measured value $x^{\rm (obs)}$ from the fit, and the true input of the simulation $x^{\rm (true)}$. The grids contain 350 nodes at which the systematic effects are evaluated. 

There is no trivial method to remove the bias on the parameters. However, assuming that the parameter space is smooth near the best fit to the observations, one could in principle evaluate the bias-free distribution from 
\begin{align}
	x^{\rm (true)} & = x^{\rm (obs)} - b(x^{\rm (true)}), \\
	b(x^{\rm (true)}) & \simeq b(x^{\rm (obs)}) + (x^{\rm (true)} - x^{\rm (obs)}) \nabla b(x),
\end{align}
which simplifies to
\begin{equation}
	x^{(true)} \simeq x^{\rm (obs)}- b(x^{\rm (obs)})/( 1 + \nabla b(x)).
\end{equation}
Here again, $x$ represents either $a_1$ or $a_3$, and $\nabla b(x)$ is the gradient estimated from the grid at the position $x^{\rm (obs)}$. However, in order to have a more robust estimate of the bias, we choose to just evaluate the average systematics $\overline{b(x^{\rm (true)})} = 1/N\,\sum^{N}_i{b(x_i^{\rm (obs)})/( 1 + \nabla_i b(x))}$ using the $N$ samples within the $2\sigma$ volume of the probability distribution of $a_1$ or $a_3$. Note that using an average over a $1\sigma$, $2\sigma$ or $3\sigma$ volume changes the average systematics by a few percent. 

Using the above-mentioned method,  we estimate that $a_1$ is likely underestimated by $5.5\%$ ($64$~nHz) and $a_3$ is overestimated by $7.5\%$ ($4$~nHz) for HD~187160. As for HD~173170, the $a_1$ might be overestimated by $2.3\%$ ($13$~nHz) and $a_3$ underestimated by $10.5\%$ ($3$~nHz). This remains small compared to the achieved precision for those stars.

\clearpage

\begin{figure*}[ht]
  \begin{center}	
  	\subfigure{\epsfig{figure=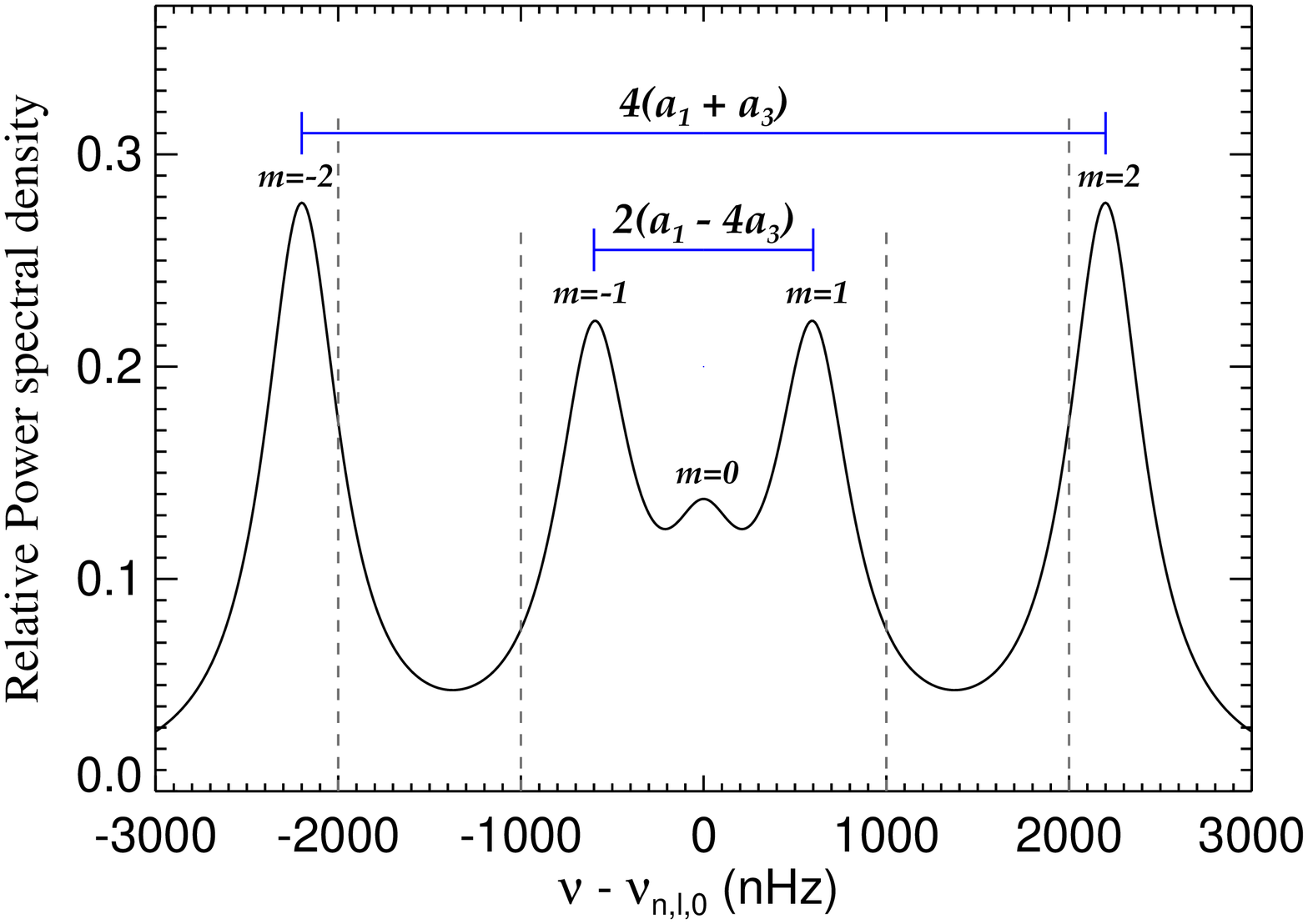, angle=0, width=12cm}}
  \end{center}
\caption{\textbf{Illustration of the power spectrum of an $l=2$ multiplet.} The stellar inclination is set to $65^\circ$ and $a_3/a_1=0.1$ (equator faster than the pole). The vertical dashed lines represent a case of a uniform latitudinal rotation ($a_3=0$).}
\label{fig:spec_l2_a1a3}
\end{figure*} 

\begin{figure*}[ht]
  \begin{center}	
  	\subfigure{\epsfig{figure=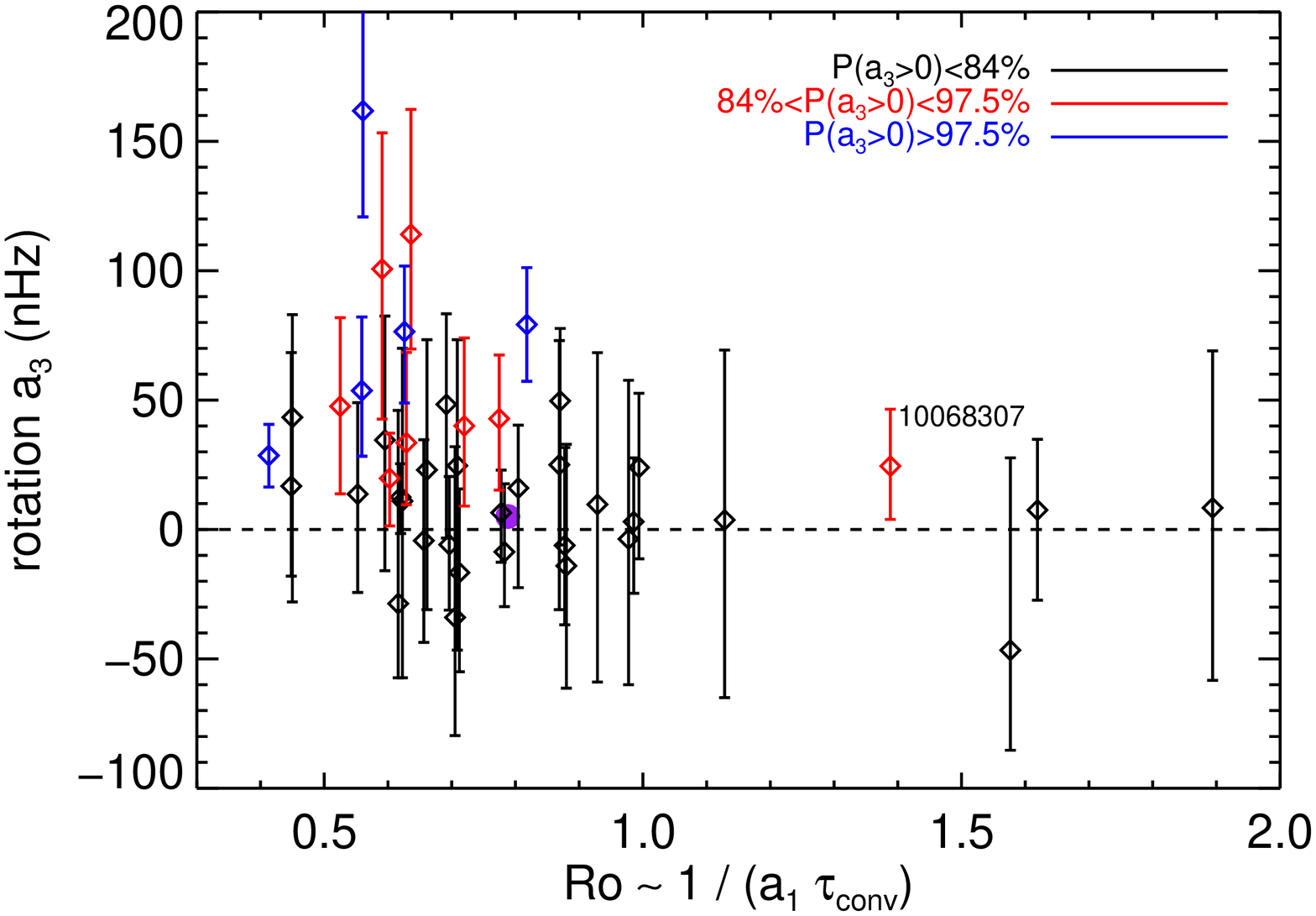, angle=0, width=12cm}}
  \end{center}
\caption{\textbf{Coefficient $a_3$ as function of the Rossby number $Ro$.} The term $\tau_{\mathrm{conv}}$ is the convective turnover time determined from stellar modeling. The solid purple circle is the Sun -- using a standard solar model, the so-called Model S \cite{JCD1996Science}. 
Most of stars with a detected solar-like differential rotation have $Ro < 1$ (except KIC 10068307). This is consistent with theoretical predictions \cite{Thompson2003}.} 
\label{fig:rossby}
\end{figure*}

\begin{figure*}[t]
  \begin{center}
	\subfigure{\epsfig{figure=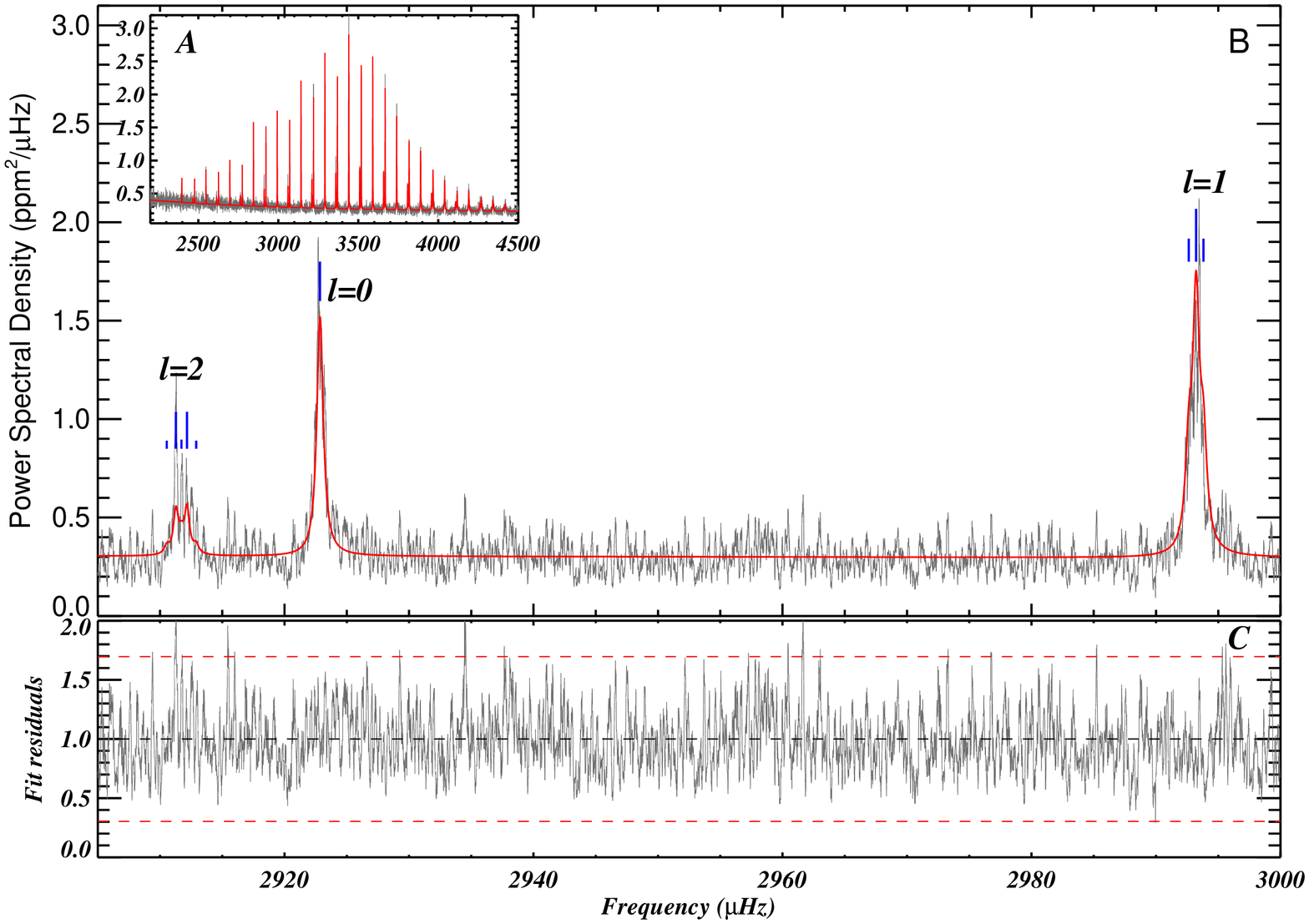, angle=0, width=16cm}}    
  \end{center}
\caption{\textbf{Spectrum for HD~173701.} \textbf{(A) }
Figure of the whole region with pulsations (gray) with the best fit model (red) superimposed. \textbf{(B)}  A zoom in the range 2905-3000 $\mu$Hz. Individual modes (blue ticks) are separated for modes of degree $l=1$ and $l=2$. The height of the ticks denotes the mode visibility at a stellar inclination of $40^\circ$. \textbf{(C)} Residual of the fit (calculated as $y(\nu)/\mathcal{M}(\nu)$) along with the $95\%$ confidence interval that the signal comes from noise (red dashed lines). Any point beyond those lines has a $5\%$ or less chance of being noise.}
\label{fig:spectrum:hd173701}
\end{figure*}

\begin{figure*}[t]
  \begin{center}
	\subfigure{\epsfig{figure=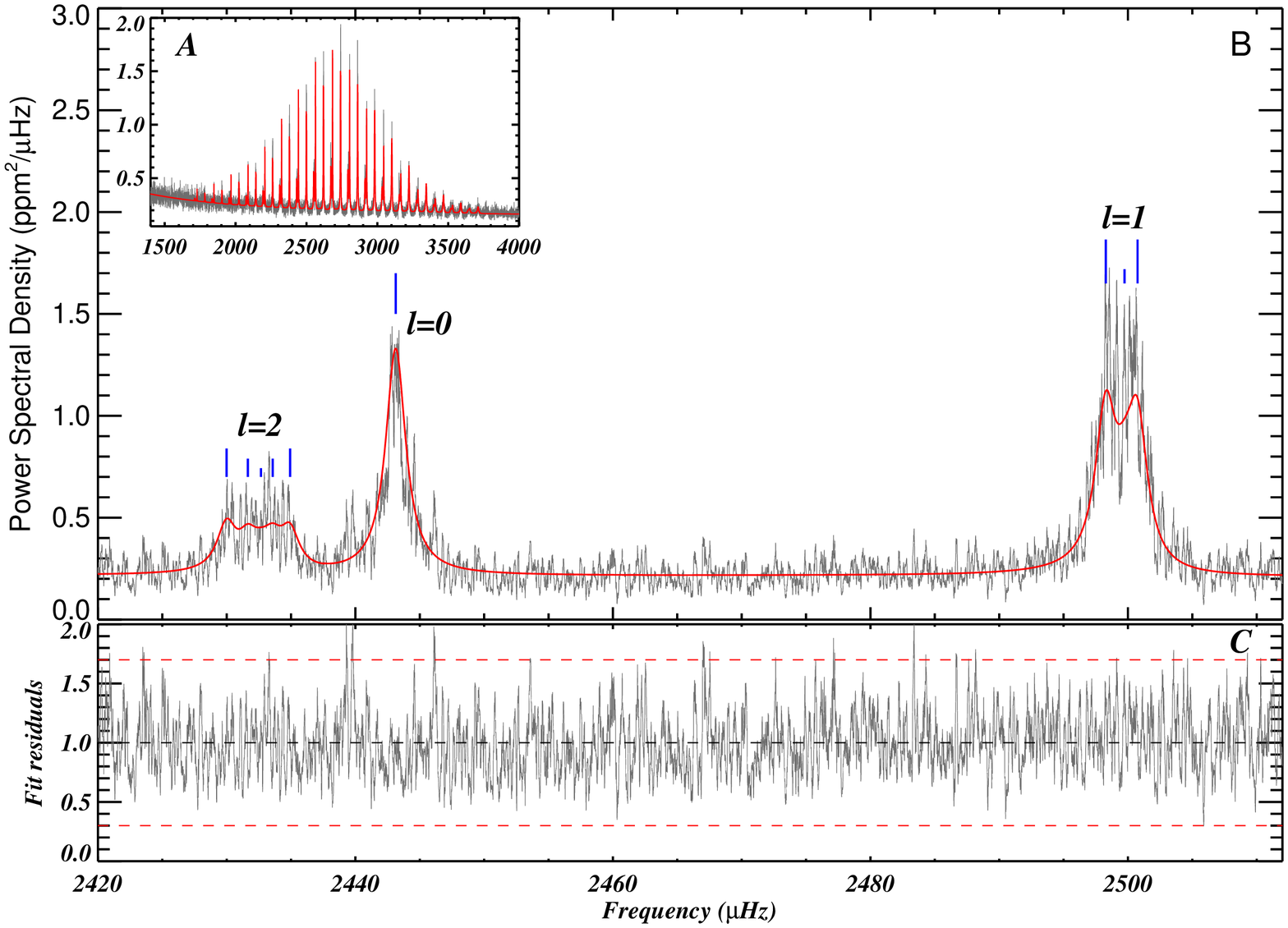, angle=0, width=16cm}} 
  \end{center}
\caption{\textbf{Spectrum for HD~187160.} The legend is similar to Figure \ref{fig:spectrum:hd173701}, but the height of the blue ticks is for a stellar inclination of $68^\circ$.}
\label{fig:spectrum:hd187160}
\end{figure*} 
 
 \begin{figure*}[ht]
  \begin{center}
	\subfigure{\epsfig{figure=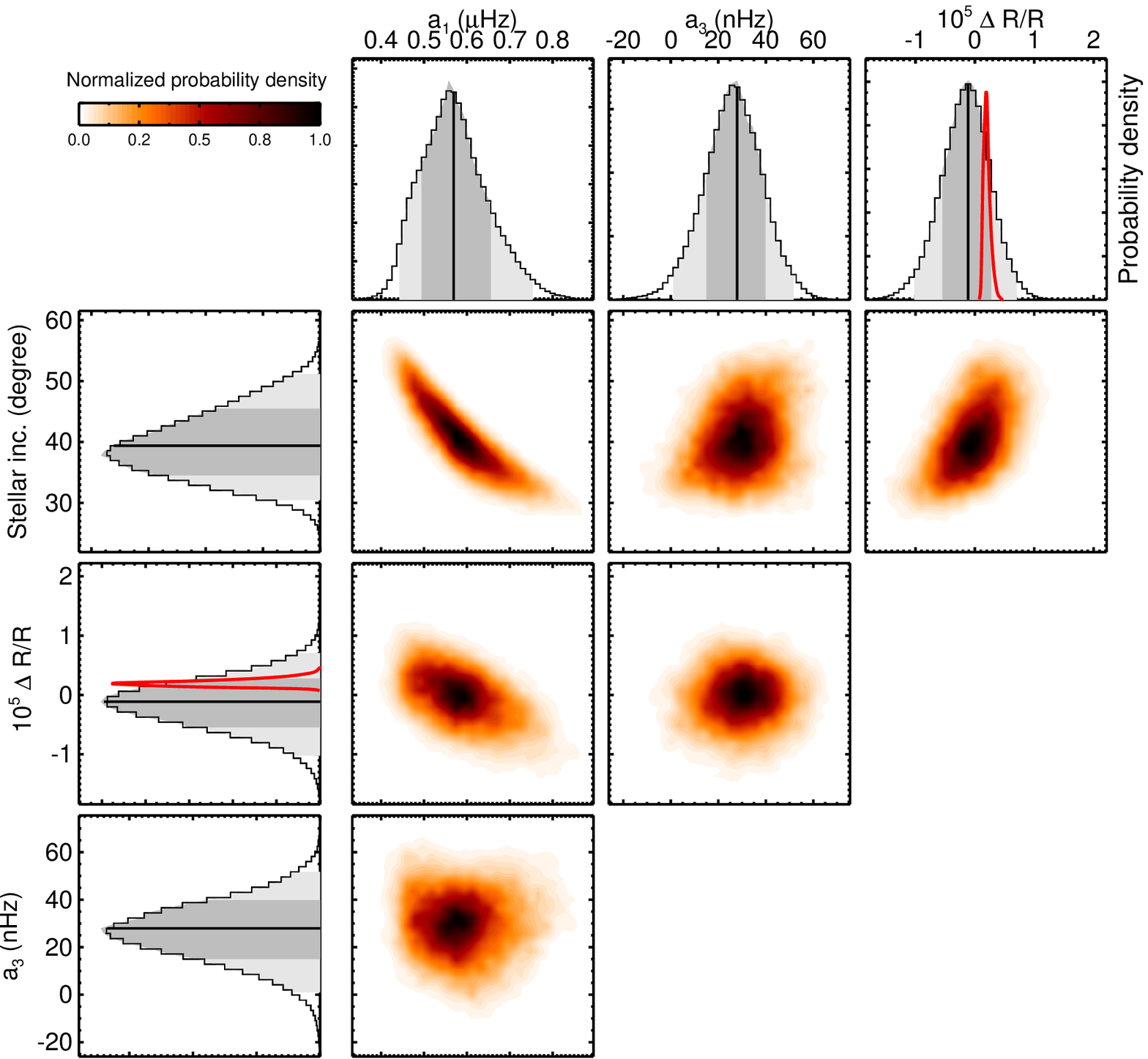, angle=0, width=14cm}}
  \end{center}
\caption{\textbf{Correlations for rotation coefficient  $a_1$ and $a_3$ for HD~173701 derived from the power spectrum.} 
The asphericity coefficient $\Delta R/R$ and the stellar inclination are also shown. The $a_3>0$ coefficient denotes that the latitudinal differential rotation is of solar-type at more than $2 \sigma$ confidence shown in light gray area. The star has a probability of $61\%$ to be prolate ($\Delta R/R<0$), but due to large uncertainties, the asphericity remains marginally consistent with a pure centrifugal distortion (red curve).}  
\label{fig:raw_results:1}
\end{figure*} 

\begin{figure*}[ht]
  \begin{center}	
\subfigure{\epsfig{figure=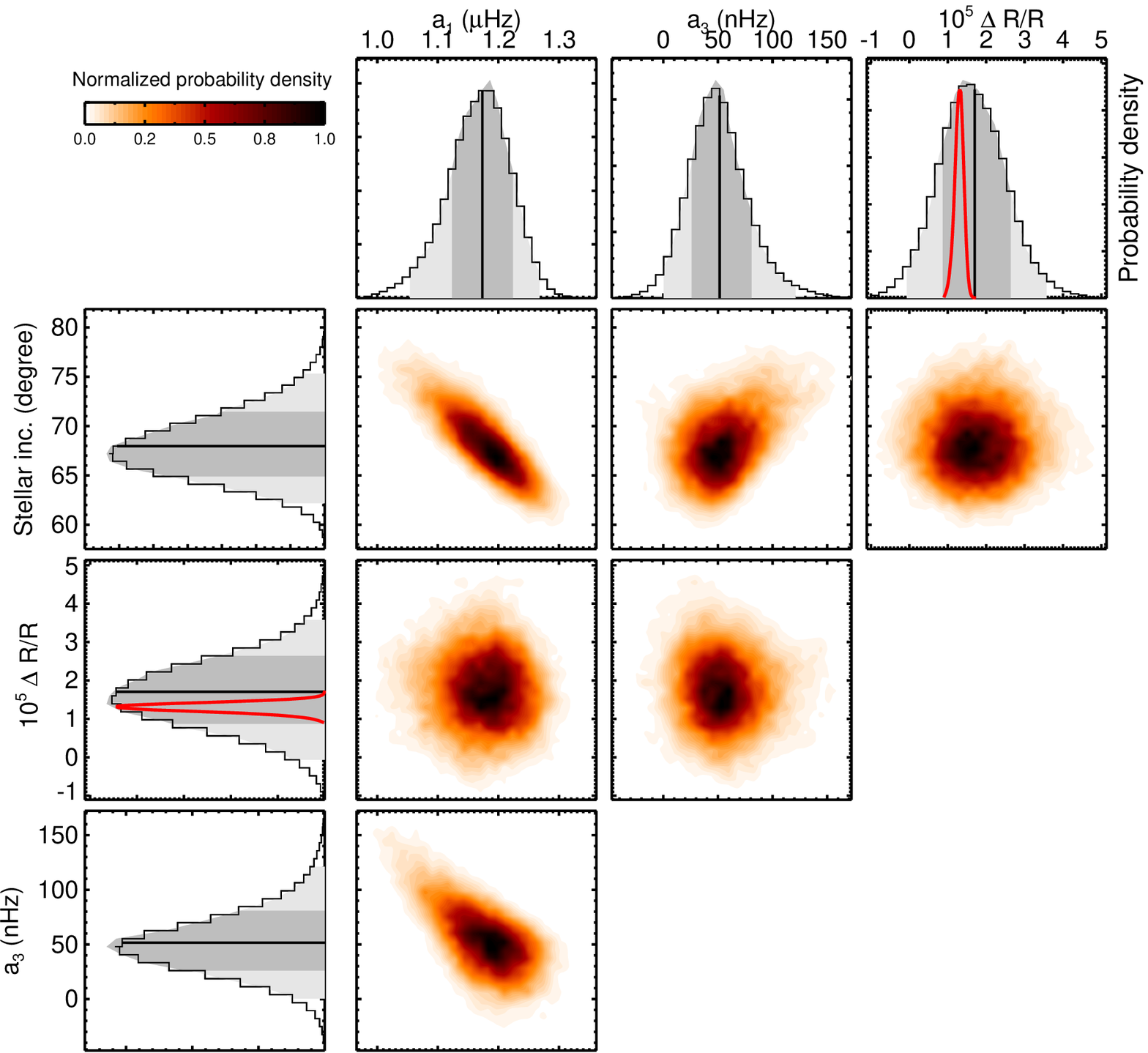, angle=0, width=14cm}}
  \end{center}
\caption{\textbf{Correlations for rotation coefficient  $a_1$ and $a_3$ for HD~187160 derived from the power spectrum.}  The legend is the same as figure \ref{fig:raw_results:1}. However, here the asphericity is consistent with a pure centrifugal distortion (red curves). The star is oblate ($\Delta R/R>0$) with probability of $98\%$.}  
\label{fig:raw_results:2}
\end{figure*} 

\begin{figure*}[ht]
  \begin{center}
	\subfigure{\epsfig{figure=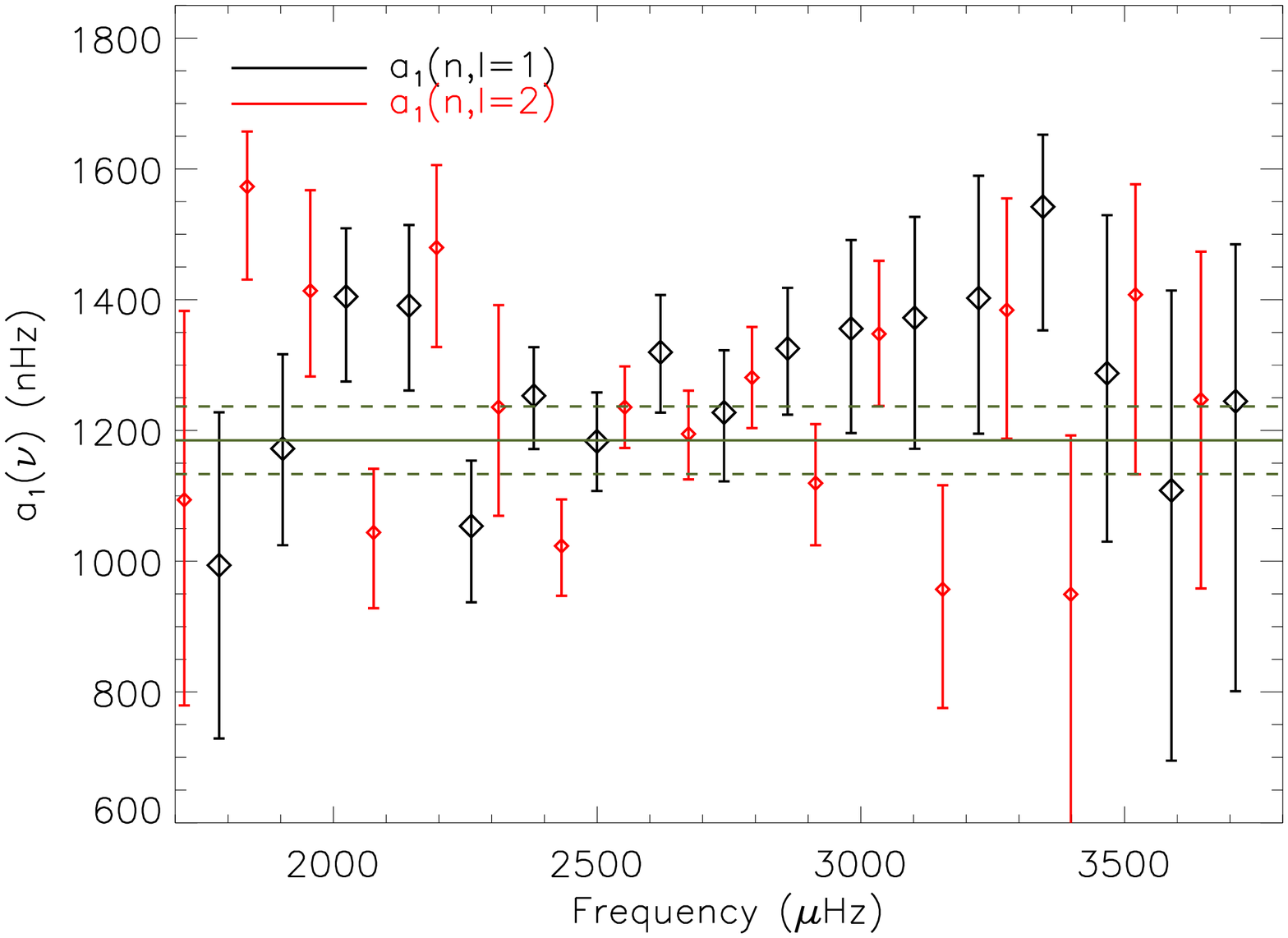, angle=0, width=12cm}}
  \end{center}
\caption{\textbf{Coefficient $a_1$ in function of the frequency for HD~187160.} Diamonds show the result of a fit of HD~187160 using independent $a_1(n,l)$ coefficients for all observed $l=1$ (black) and $l=2$ (red) modes, compared to the best fitting model of a constant $a_1$ (solid line) and its confidence interval at $1\sigma$ (dashed lines). Fitting individual splittings gives large uncertainties. All measures of $a_1(n,l)$ remain consistent at $2\sigma$ with the fit of the power spectrum with a constant $a_1$, such that there is no evidence of a significant radial differential rotation in the stellar interior.}
\label{fig:a1n}
\end{figure*} 

\begin{figure*}[ht]
  \begin{center}
	\subfigure{\epsfig{figure=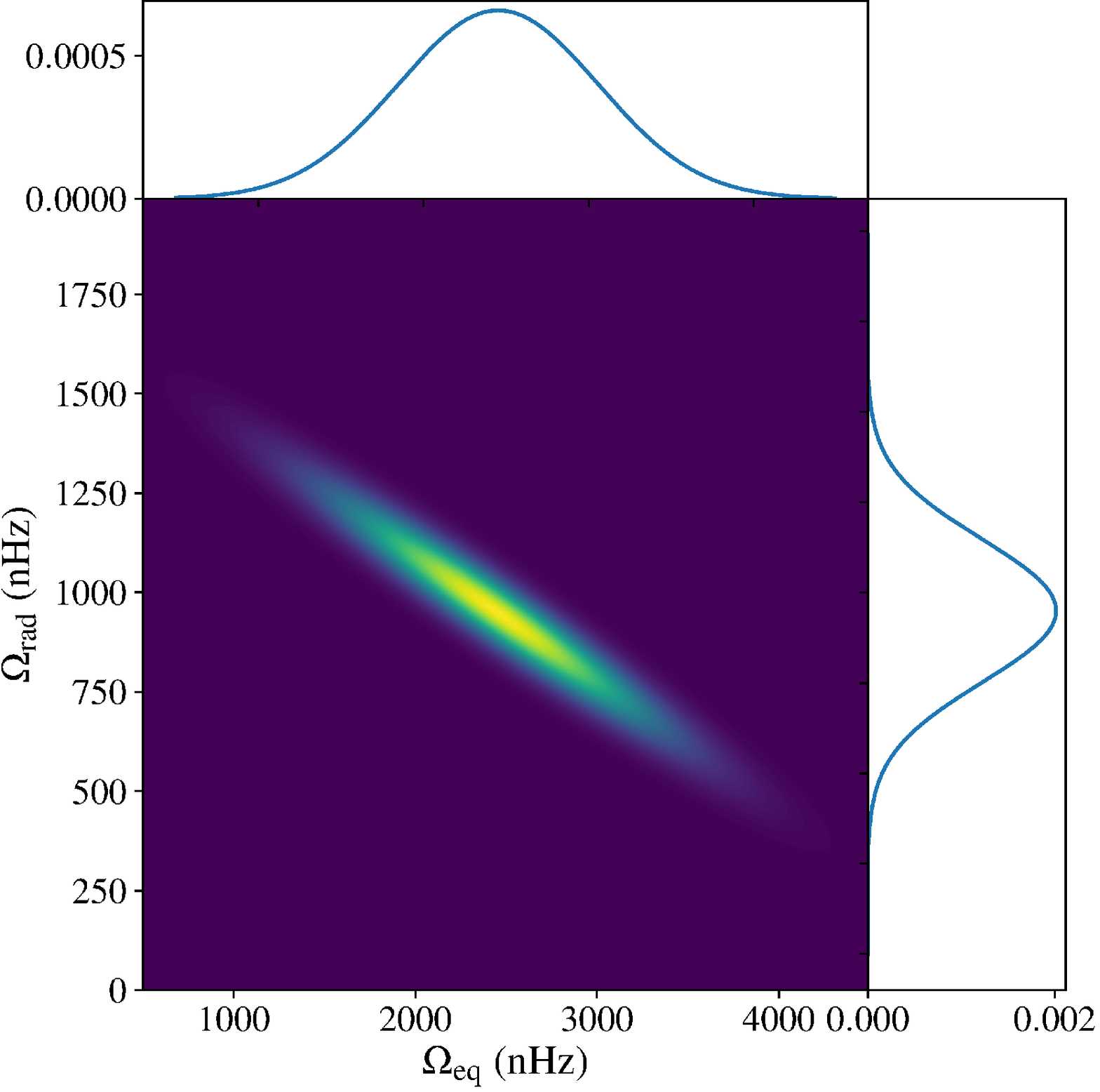, angle=270, width=10cm}}
  \end{center}
\caption{\textbf{Rotation coefficients for a cylindrical rotation.} Posterior distribution for $\Omega_{\rm rad}$ and $\Omega_{\mathrm{eq}}$  of HD~187160 for a tangent-cylinder rotation (no tachoclyne) in the convective zone.} 
\label{fig:rot:cyl1}
\end{figure*}


\clearpage

\begin{table*}
\caption{\textbf{Derived $a_1$ and $a_3$ from power spectrum fitting.} Asteroseismic measurements of $a_1$ and $a_3$, along with the probability of having $a_3$ positive (solar-like rotation). The stars are identified using the KIC or HD number, and coordinates (R.A., Dec.). Fundamental parameters of these stars are in \cite{Aguirre2017} and in Table S2.}
\centering \footnotesize   
\begin{tabular}{c|c|c|c|c|c|c}
KIC               &    HD        &  R.A. (deg)    & Dec. (deg)  & $a_1$  (nHz)         &       $a_3$ (nHz)  &    $P(a_3>0)$ (\%)  \\ \hline\hline
003427720   &   ------       &  286.4386   &   38.52293 &  $337.7^{+354.1}_{-150.0}$  &   $-46.7^{+74.3}_{-38.7}$  &   $24.730$ \\
007871531   &   ------       &   282.9151   &   43.63609  & $422.5^{+42.6}_{-60.3}$  &   $-28.6^{+74.6}_{-28.7}$  &   $27.041$ \\
009410862   &   ------       &  294.4862   &   45.96451  &  $683.1^{+566.1}_{-324.7}$  &   $-34.0^{+66.0}_{-45.7}$  &   $29.081$ \\
009098294   &   ------       &  295.0883   &   45.48915  &  $473.0^{+148.0}_{-100.9}$  &   $-16.7^{+32.3}_{-38.3}$  &   $29.625$ \\
007970740   &  186306   &  295.4402   &   43.74764  &  $321.8^{+49.2}_{-47.4}$  &   $-8.7^{+26.3}_{-21.2}$  &   $34.719$ \\
010454113   &   176153  &  284.1526   &   47.65641  &  $1351.7^{+307.4}_{-247.5}$  &   $-14.1^{+47.0}_{-47.3}$  &   $37.973$ \\
006106415   &   177153  &  285.4153   &   41.49009 &  $722.6^{+37.5}_{-41.1}$  &   $-5.9^{+26.3}_{-25.3}$  &   $40.266$ \\
005950854   &   ------       &  288.7837   &   41.29874  &  $375.0^{+1295.8}_{-254.2}$  &   $-6.2^{+37.8}_{-30.7}$  &   $43.549$ \\
012009504   &  234856   &  289.4409   &   50.48006  &   $1211.2^{+64.4}_{-73.1}$  &   $-4.3^{+39.0}_{-39.3}$  &   $44.349$ \\
008938364   &   ------       &  285.5413   &   45.27761  &  $229.1^{+741.5}_{-141.6}$  &   $-3.7^{+61.3}_{-56.3}$  &   $47.062$ \\
007106245   &   ------       &  288.0094   &   42.67186  &  $779.6^{+682.9}_{-506.3}$  &   $3.7^{+65.7}_{-68.7}$  &   $51.255$ \\
003656476   &   ------       &  294.2033   &   38.71578  &  $239.2^{+71.5}_{-40.9}$  &   $3.0^{+24.7}_{-27.7}$  &   $53.831$ \\
008150065   &   ------       &  285.0289   &   44.02919  &  $649.3^{+499.4}_{-266.4}$  &   $8.3^{+60.7}_{-66.7}$  &   $53.972$ \\
008760414   &   ------       &  294.4768   &   44.98445  &  $430.0^{+622.5}_{-317.5}$  &   $9.7^{+58.7}_{-68.7}$  &   $54.836$ \\
010079226   &   ------       &  293.4699   &   47.04572  &  $672.9^{+205.2}_{-101.0}$  &   $11.0^{+59.0}_{-68.3}$  &   $55.696$ \\
006933899   &   ------       &  286.7431   &   42.43562  &  $333.3^{+40.7}_{-39.0}$  &   $7.5^{+27.3}_{-34.8}$  &   $58.149$ \\
004914923   &   ------       &  289.1454   &   40.04731  &  $564.0^{+97.4}_{-109.1}$  &   $6.5^{+16.5}_{-19.1}$  &   $62.979$ \\
008394589   &   190166  &  300.4643   &   44.35391  &  $1066.0^{+60.8}_{-69.9}$  &   $13.7^{+35.3}_{-38.0}$  &   $63.649$ \\
008424992   &   ------       &  289.6384   &   44.40449  &  $370.8^{+479.1}_{-204.2}$  &   $24.7^{+48.7}_{-71.3}$  &   $64.311$ \\
006603624   &   ------       &  291.0467   &   42.05270  &  $291.4^{+1082.4}_{-183.2}$  &   $16.0^{+24.3}_{-38.5}$  &   $66.281$ \\
011772920   &   ------       &  296.4415   &   49.98858  &  $314.9^{+59.7}_{-61.9}$  &   $23.0^{+50.3}_{-54.0}$  &   $66.696$ \\
010516096   &   ------       &  282.3902   &   47.71106  &  $483.1^{+64.1}_{-62.1}$  &   $25.0^{+48.0}_{-56.0}$  &   $68.502$ \\
007296438   &   ------       & 295.8723   &   42.88115  &  $646.7^{+360.0}_{-463.3}$  &   $43.3^{+39.7}_{-71.3}$  &   $73.715$ \\
008179536   &   ------       &  296.3499   &   44.07664  &  $1984.5^{+167.9}_{-211.8}$  &   $34.5^{+48.0}_{-50.5}$  &   $74.281$ \\
008228742   &   ------       &  290.1997   &   44.15524  &  $903.4^{+227.6}_{-349.2}$  &   $24.0^{+28.7}_{-35.3}$  &   $75.867$ \\
003735871   &   ------       &  288.0693   &   38.81768  &  $828.5^{+221.4}_{-170.5}$  &   $48.3^{+35.0}_{-51.7}$  &   $82.249$ \\
007680114   &   ------       &  290.985    &   43.31459  &  $482.8^{+182.2}_{-184.5}$  &   $49.7^{+28.0}_{-55.7}$  &   $82.305$ \\
010963065   &   176693  &  284.7862   &   48.42323  &  $1011.4^{+95.3}_{-98.0}$  &   $19.7^{+17.5}_{-18.4}$  &   $84.451$ \\
010068307   &   ------       &  288.9779   &   47.06124  &  $866.8^{+111.9}_{-154.0}$  &   $24.5^{+22.0}_{-20.6}$  &   $87.199$ \\
005184732   &   182756  &  291.1264   &   40.32326  &  $552.1^{+60.0}_{-55.8}$  &   $40.0^{+34.0}_{-31.0}$  &   $88.164$ \\
009965715   &   ------       &  297.3332   &   46.82664  &  $1967.2^{+83.5}_{-88.7}$  &   $47.6^{+34.3}_{-33.8}$  &   $90.895$ \\
009955598   &   ------       &  293.6792   &   46.85276  &  $382.9^{+80.1}_{-84.5}$  &   $33.5^{+34.9}_{-23.9}$  &   $90.901$ \\
012258514   &   183298  &  291.592    &   50.98724  &  $815.7^{+187.5}_{-287.9}$  &   $42.8^{+24.6}_{-27.6}$  &   $92.285$ \\
009025370   &   184401  &  293.0609   &   45.31069  &  $469.1^{+86.8}_{-72.0}$  &   $100.6^{+52.6}_{-58.0}$  &   $93.735$ \\
009139151   &   ------       &  284.0886   &   45.51475  &  $974.5^{+97.5}_{-96.5}$  &   $114.0^{+48.3}_{-44.3}$  &   $97.446$ \\
008379927   &   187160  &  296.6721   &   44.34853  &  $1179.7^{+49.0}_{-56.9}$  &   $53.6^{+28.5}_{-25.3}$  &   $97.977$ \\
008006161   &   173701  &  281.1465   &   43.83331  &  $561.1^{+70.9}_{-67.4}$  &   $28.6^{+12.0}_{-12.2}$  &   $98.443$ \\
008694723   &   ------       &  293.9608   &   44.88052  &  $1679.2^{+111.7}_{-122.1}$  &   $76.5^{+25.3}_{-27.6}$  &   $99.645$ \\
006225718   &   187637  &  297.3256   &   41.58246  &  $1141.3^{+225.7}_{-152.8}$  &   $79.2^{+22.0}_{-22.0}$  &   $99.930$ \\
007510397   &   177412  &  285.6642   &   43.11748 &  $1532.9^{+158.6}_{-126.4}$  &   $161.8^{+42.0}_{-41.0}$  &   $99.980$ \\
\end{tabular}
\end{table*}

\begin{table*}
\caption{\textbf{Properties of the stellar models used for the inversion of the rotation profile.} Only stars with probability of having $a_3$ positive greater than $84\%$ are shown (see also Table S1). The parameters are computed using the \textsc{astfit} pipeline from \cite{Aguirre2017}. $\teff$, Z/X, $\mathrm{M/M_{\odot}}$, $\mathrm{R/R_{\odot}}$ are the effective temperature, fraction of heavy elements, the mass and radius of the stars, respectively. $\mathrm{D_{CZ}}$ is the thickness of the convective zone.}
\centering
\begin{tabular}{ccccccc}
         KIC  &   $\teff$  &       Z/X  &  $\mathrm{M/M_{\odot}}$  &  $\mathrm{R/R_{\odot}}$  &   Age (Gyr)  &    $\mathrm{D_{CZ}}$    \\ \hline\hline
     5184732  &      5896  &   0.05342  &           1.21  &          1.344  &        4.39  &       0.259      \\ 
     6225718  &      6299  &   0.02164  &           1.15  &          1.233  &        3.03  &       0.186      \\ 
     7510397  &      6148  &   0.04257  &           1.45  &          1.906  &        2.69  &       0.192      \\ 
     8006161  &      5445  &   0.04967  &           0.95  &          0.917  &        4.49  &       0.313      \\ 
     8694723  &      6105  &   0.01719  &           1.11  &          1.536  &        5.45  &       0.180      \\ 
     9025370  &      5485  &   0.03051  &           0.97  &          1.000  &        5.24  &       0.289      \\ 
     9139151  &      6319  &   0.02721  &           1.15  &          1.151  &        2.74  &       0.225      \\ 
     9955598  &      5431  &   0.02721  &           0.90  &          0.886  &        6.64  &       0.313      \\ 
     9965715  &      6157  &   0.01737  &           1.08  &          1.277  &        4.27  &       0.174      \\ 
    10068307  &      6404  &   0.04778  &           1.61  &          2.165  &        2.02  &       0.166      \\ 
    10963065  &      6147  &   0.02202  &           1.08  &          1.229  &        4.30  &       0.226      \\ 
    12258514  &      5918  &   0.02706  &           1.14  &          1.550  &        5.78  &       0.216      \\ 
     8379927  &      6184  &   0.02461  &           1.12  &          1.123  &        1.87  &       0.210      \\ \hline
\end{tabular}
\end{table*}

\begin{table*}
\caption{\textbf{Results of the rotation inversion.} Only stars with a significant ($>1\sigma$) latitudinal differential rotation are shown. $\Omega_0$ and $\Omega_1$ are the parameters of the solar-like rotation profile (Eq.~\ref{eq:domega0}-\ref{eq:domega1}) determined by seismic inversion. $\Delta\Omega_{45^\circ}/\Omega_{\mathrm{eq}}$ is the shear factor between the equator and $45^\circ$ of latitude, and $\Delta\Omega_{90^\circ}/\Omega_{\mathrm{eq}}$ is the shear factor between the equator and the pole.} 
\centering
\begin{tabular}{ccccc}
         KIC  &   $\Omega_0 / 2 \pi$  (nHz) &      $\Omega_1 / 2 \pi$ (nHz) &   $\Delta\Omega_{45^\circ}/\Omega_{\mathrm{eq}}$ &   $\Delta\Omega_{90^\circ}/\Omega_{\mathrm{eq}}$ \\ \hline\hline
     5184732  &              $560 \pm 61$  &                $-150 \pm 143$  &                   $-0.74_{0.43}^{0.53}$  &                   $-1.47_{0.85}^{1.05}$ \\ 
     6225718  &            $1179 \pm 192$  &                $-364 \pm 104$  &                   $-0.79_{0.18}^{0.19}$  &                   $-1.58_{0.37}^{0.37}$ \\ 
     7510397  &            $1590 \pm 163$  &                $-776 \pm 204$  &                   $-1.06_{0.15}^{0.18}$  &                   $-2.12_{0.29}^{0.36}$ \\     
     8006161  &              $566 \pm 69$  &                 $-104 \pm 45$  &                   $-0.54_{0.18}^{0.20}$  &                   $-1.08_{0.35}^{0.40}$ \\ 
     8379927  &             $1188 \pm 53$  &                $-241 \pm 118$  &                   $-0.57_{0.21}^{0.22}$  &                   $-1.13_{0.43}^{0.44}$ \\ 
     8694723  &            $1715 \pm 127$  &                $-374 \pm 131$  &                   $-0.62_{0.14}^{0.18}$  &                   $-1.25_{0.28}^{0.37}$ \\ 
     9025370  &              $481 \pm 94$  &                $-356 \pm 244$  &                   $-1.40_{0.30}^{0.55}$  &                   $-2.81_{0.61}^{1.09}$ \\ 
     9139151  &              $982 \pm 94$  &                $-474 \pm 221$  &                   $-1.06_{0.27}^{0.32}$  &                   $-2.12_{0.54}^{0.63}$ \\ 
     9955598  &              $383 \pm 79$  &                $-134 \pm 109$  &                   $-0.79_{0.58}^{0.49}$  &                   $-1.57_{1.16}^{0.98}$ \\ 
     9965715  &             $1982 \pm 89$  &                $-226 \pm 166$  &                   $-0.37_{0.21}^{0.25}$  &                   $-0.73_{0.41}^{0.49}$ \\ 
    10068307  &             $875 \pm 136$  &                $-131 \pm 118$  &                   $-0.44_{0.30}^{0.35}$  &                   $-0.89_{0.60}^{0.71}$ \\ 
    10963065  &            $1018 \pm 103$  &                  $-81 \pm 77$  &                   $-0.27_{0.20}^{0.25}$  &                   $-0.55_{0.40}^{0.50}$ \\ 
    12258514  &             $805 \pm 235$  &                $-188 \pm 131$  &                   $-0.66_{0.25}^{0.34}$  &                   $-1.33_{0.50}^{0.68}$ \\ 
    $8379927$  &             $1122 \pm 53$  &                $-245 \pm 130$  &                   $-0.56_{0.20}^{0.25}$  &                   $1.10_{0.44}^{0.47}$ \\ \hline
\end{tabular}
\end{table*}

\begin{table*}
\caption{\textbf{Characteristics of the data retained for the analysis.} This comprises the data version available at \url{kasoc.phys.au.dk}, the used quarter interval, the total observation duration and the frequency resolution of the spectrum. The power spectra used in this work are available at \url{https://doi.org/10.7910/DVN/8SK6OL}.}
\centering \footnotesize   
\begin{tabular}{c|c|c|c|c}
      KIC        &   Version      &        Quarters    & Obs. duration (days) &  freq. resol. (nHz) \\ \hline\hline
003427720        &           1    &      Q5.1-Q17.2    &       1147.53   &      10.1  \\
003656476        &           1    &      Q5.1-Q17.2    &       1147.53   &      10.1  \\
003735871        &           1    &      Q5.1-Q17.2    &       1147.53   &      10.1  \\
004914923        &           1    &      Q5.1-Q17.2    &       1147.53   &      10.1  \\
005184732        &           1    &      Q7.1-Q17.2    &       960.85    &      12.0  \\
005950854        &           1    &      Q5.1-Q10.3    &       556.80    &      20.8  \\
006106415        &           2    &      Q6.1-Q16.3    &       1018.53   &      11.4  \\
006225718        &           1    &      Q6.1-Q17.2    &       1051.57   &      11.0  \\
006603624        &           1    &      Q5.1-Q17.2    &       1147.53   &      10.1  \\
006933899        &           1    &      Q5.1-Q17.2    &       1147.53   &      10.1  \\
007106245        &           1    &      Q5.1-Q15.3    &       1027.67   &      11.3  \\
007296438        &           3    &      Q7.1-Q11.3    &       468.17    &      24.7  \\
007510397        &           1    &      Q7.1-Q17.2    &       960.84    &      12.0  \\
007680114        &           1    &      Q5.1-Q17.2    &       1147.53   &      10.1  \\
007871531        &           1    &      Q5.1-Q17.2    &       1147.53   &      10.1  \\
007970740        &           1    &      Q6.1-Q17.2    &       1051.57   &      11.0  \\
008006161        &           1    &      Q5.1-Q17.2    &       1147.53   &      10.1  \\
008150065        &           1    &      Q5.1-Q10.3    &       556.80    &      20.8  \\
008179536        &           1    &      Q5.1-Q11.3    &       654.86    &      17.7  \\
008228742        &           1    &      Q5.1-Q17.2    &       1147.53   &      10.1  \\
008379927        &           2    &      Q2.1-Q17.2    &       1421.50   &      8.1  \\
008394589        &           1    &      Q5.1-Q17.2    &       1147.53   &      10.1  \\
008424992        &           1    &      Q7.1-Q10.3    &       370.11    &      31.3  \\
008694723        &           1    &      Q5.1-Q17.2    &       1147.53   &      10.1  \\
008760414        &           1    &      Q5.1-Q17.2    &       1147.53   &      10.1  \\
008938364        &           1    &      Q6.1-Q17.2    &       1051.57   &      11.0  \\
009025370        &           1    &      Q5.1-Q17.2    &       1147.53   &      10.1  \\
009098294        &           1    &      Q5.1-Q17.2    &       1147.53   &      10.1  \\
009139151        &           1    &      Q5.1-Q17.2    &       1147.53   &      10.1  \\
009410862        &           1    &      Q5.1-Q15.3    &       1027.67   &      11.3  \\
009955598        &           5    &      Q5.1-Q17.2    &       1147.53   &      10.1  \\
009965715        &           1    &      Q5.1-Q13.3    &       829.59    &      13.9  \\
010068307        &           1    &      Q7.1-Q17.2    &       960.85    &      12.0  \\
010079226        &           1    &      Q7.1-Q10.3    &       370.11    &      31.3  \\
010454113        &           1    &      Q5.1-Q17.2    &       1147.53   &      10.1  \\
010516096        &           1    &      Q5.1-Q17.2    &       1147.53   &      10.1  \\
010963065        &           5    &      Q2.3-Q17.2    &       1359.65   &      8.5  \\
011772920        &           1    &      Q5.1-Q17.2    &       1147.53   &      10.1  \\
012009504        &           1    &      Q5.1-Q17.2    &       1147.53   &      10.1  \\
012258514        &           1    &      Q5.1-Q17.2    &       1147.53   &      10.1  \\ \hline
\end{tabular}  
\end{table*}

\end{document}